\documentclass[pdflatex,sn-mathphys-num]{sn-jnl}

\usepackage{graphicx}%
\usepackage{subfigure}%
\usepackage{multirow}%
\usepackage{amsmath,amssymb,amsfonts}%
\usepackage{amsthm}%
\usepackage{mathrsfs}%
\usepackage[title]{appendix}%
\usepackage{xcolor}%
\usepackage{textcomp}%
\usepackage{manyfoot}%
\usepackage{booktabs}%
\usepackage{algorithm}%
\usepackage{algorithmicx}%
\usepackage{algpseudocode}%
\usepackage{listings}%
\usepackage[utf8]{inputenc}
\usepackage[T1]{fontenc}
\usepackage{booktabs}%
\usepackage{subcaption}%
\usepackage{svg}%

\definecolor{rowgray}{HTML}{F5F5F5}
\usepackage[most]{tcolorbox}      

\definecolor{pastelblue}{RGB}{220,230,241}

\tcbset{
  mytablebox/.style={
    colback=pastelblue,   
    boxrule=0pt,          
    arc=0pt,              
    left=4pt, right=4pt, top=4pt, bottom=4pt
  }
}

\usepackage{adjustbox}
\usepackage{array}

\newcolumntype{R}[2]{%
    >{\adjustbox{angle=#1,lap=\width-(#2)}\bgroup}%
    l%
    <{\egroup}%
}
\newcommand*\rot{\multicolumn{1}{R{45}{1em}}}

\theoremstyle{thmstyleone}%
\theoremstyle{thmstyletwo}%

\theoremstyle{thmstylethree}%
\newcommand{\rotvert}{\rotatebox[origin=c]{90}{$\vert$}}

\begin{document}

\title[Article Title]{
Harnessing the Full Potential of RRAMs through Scalable and Distributed In-Memory Computing with Integrated Error Correction}


\author[1]{\fnm{Huynh Q. N.} \sur{Vo}}\email{lucius.vo@okstate.edu}

\author[2]{\fnm{Md Tawsif Rahman} \sur{Chowdhury}}\email{mtawsifrc@wayne.edu}

\author*[1]{\fnm{Paritosh} \sur{Ramanan}}\email{paritosh.ramanan@okstate.edu}

\author*[3]{\fnm{Murat} \sur{Yildirim}}\email{murat@wayne.edu}

\author*[2]{\fnm{Gozde} \sur{Tutuncuoglu}}\email{gozde@wayne.edu}

\affil*[1]{\orgdiv{School of Industrial Engineering and Management}, \orgname{Oklahoma State University}, \orgaddress{\city{Stillwater}, \postcode{74078}, \state{OK}, \country{USA}}}

\affil[2]{\orgdiv{Department of Electrical and Computer Engineering}, \orgname{Wayne State University}, \orgaddress{\city{Detroit}, \postcode{48202}, \state{MI}, \country{USA}}}

\affil[3]{\orgdiv{Department of Industrial and System Engineering}, \orgname{Wayne State University}, \orgaddress{\city{Detroit}, \postcode{48202}, \state{MI}, \country{USA}}}


\abstract{
Exponential growth in global computing demand is exacerbated due to the higher-energy requirements of conventional architectures, primarily due to energy-intensive data movement. In-memory computing with Resistive Random Access Memory (RRAM) addresses this by co-integrating memory and processing, but faces significant hurdles related to device-level non-idealities and poor scalability for large computing tasks. Here, we introduce \textbf{MELISO+} (In-\textbf{Me}mory \textbf{Li}near \textbf{So}lver), a full-stack, distributed framework for energy-efficient in-memory computing. MELISO+ proposes a novel two-tier error correction mechanism to mitigate device non-idealities, and develops a distributed RRAM computing framework to enable matrix computations exceeding dimensions of 65,000×65,000. This approach reduces first- and second-order
arithmetic errors due to device non-idealities by over 90\%, enhances energy efficiency by three to five
orders of magnitude, and decreases latency 100-fold. Hence, MELISO+ allows lower-precision RRAM devices to outperform high-precision device alternatives in accuracy, energy and latency metrics. By unifying algorithm-hardware co-design with scalable architecture, MELISO+ significantly advances sustainable, high-dimensional computing suitable for applications like large language models and generative AI.}



\maketitle
\thispagestyle{empty}

\section{Introduction}\label{sec:Introduction}

Energy demands for computing have been increasing at an unprecedented rate in recent years, with projected consumption anticipated to surpass today’s global power demand by 2030~\cite{challe6010117}. Conventional von Neumann computing systems cannot sustainably address this exponential growth due to inherent challenges, such as the issues arising from the spatial separation of computation and memory functionalities. For instance, a recent study of Google server workloads revealed that 62.7\% of total energy consumption was attributed to data movement rather than actual computation~\cite{10.1145/3296957.3173177}. In-memory computing emerges as a promising technology to address this challenge by co-locating memory and computation and enabling reductions of 2-3 orders of magnitude in energy consumption and 20-fold improvements in latency~\cite{Khan2024TheLO}. 
Within in-memory computing enabling technologies, RRAM devices are particularly popular as they can store up to $2048$ conductance levels with sub-nanosecond latency and picojoule-scale write energy~\cite{ rao_thousands_2023}. However, widespread adoption of RRAM device technology is hindered by two key challenges: variability and scalability. Variability arises due to inherent device non-idealities in RRAMs, which manifest themselves as additional error terms causing significant inaccuracies in computing tasks~\cite{10766540}. Scalability is another challenge, caused by limitations on the maximum feasible size of RRAM arrays due to physical constraints such as sneak paths and parasitic interconnect resistances~\cite{https://doi.org/10.1002/admt.202400585,ielmini_resistive_2025}. Addressing these key issues can pave the way for large-scale adoption of sustainable computing enabled by RRAM devices.

Existing studies address errors caused by non-idealities in RRAM operations by proposing a range of error detection and correction methods, 
each with a different set of trade-offs. The first set of methods builds on traditional parity checks that encode redundant information to detect errors, which incurs 
substantial burdens in latency, energy consumption, and circuit footprint~\cite{10088479, 10.1145/3489517.3530526,9107606, 10.1145/3386360, 10616083}. The second set of methods, called arithmetic logic schemes, applies additional linear transformations to verify whether values are integer multiples of a predefined constant, enabling error detection through structured algebraic constraints. The applicability of these methods is limited only to a subset of computational tasks, and they also lack the means to correct the detected errors.
\cite{1671530, 9420944, 10567894}. Architectural strategies—e.g., redundancy and remapping—improve fault tolerance at the cost of increasing RRAM array size and lack the capability to address transient errors~\cite{10567894, 10616063, 8326998, 5695530}. Algorithmic methods adapt the computation method itself to tolerate errors from RRAM encoding, rather than correcting device output errors. This approach is inherently limited to a small subset of computational tasks.~\cite{8060459, 8806787, 9586269,10242326}.

Significant latency, overhead, and limited generalizability of the proposed error correction (EC) methods will become more critical as problem sizes grow, further constraining scalability and efficiency. In fact, most existing RRAM computational applications are only demonstrated in small-scale use cases \cite{8558705, doi:10.1073/pnas.1815682116, Mahmoodi_Prezioso_Strukov_2019, 7409720, 8106770}. These limitations underscore the need for unified frameworks that are not only robust to hardware variability but also capable of supporting large-scale and diverse computational tasks. This capability is critical for using RRAM systems for modern computing tasks such as constrained optimization, large language models, and generative artificial intelligence applications \cite{Sengar_Hasan_Kumar_Carroll_2024}. To meet this need, our work expands beyond isolated EC methods by introducing a set of computational methods designed specifically for RRAM systems, coupled with a scalable, distributed, and general-purpose simulation platform that integrates novel EC techniques with a full-stack virtualization and benchmarking pipeline designed specifically for RRAM-based in-memory computing.

In this paper, we propose MELISO+, a suite of RRAM computational methods and a full-stack simulation/benchmarking framework designed for accurate, energy-efficient, and large-scale matrix-vector computations. MELISO+ is a first-of-its-kind framework that jointly addresses two key challenges: 
(i) scalability limitations were resolved through the addition of parallel distributed computing and virtualization capabilities that can seamlessly break down computation tasks that can be performed by tiles of RRAM arrays, and (ii) variability issues were addressed through a novel EC algorithm that was developed specifically for RRAM devices and implemented within the MELISO+ framework. MELISO+ introduces a distributed computation environment that leverages memory crossbar arrays (MCAs) operated through the Message Passing Interface (MPI), enabling scalable and parallelizable execution across systems with constrained hardware resources. Our virtualization mechanism allows matrices of dimensions exceeding $65{,}000 \times 65{,}000$ to be encoded and processed on fixed-size MCAs, supporting diverse computational workloads without hardware reconfiguration. On the algorithmic side, MELISO+ develops an integrated incremental weight updating mechanism and an EC algorithm that is tailored specifically for RRAM devices. The proposed EC model reduces first- and second-order
arithmetic errors caused by device non-idealities by over 90\%, enhances energy efficiency by three to five
orders of magnitude, and decreases latency 100-fold. These techniques also allow lower performing devices (e.g., AlO\textsubscript{x}-HfO\textsubscript{2}~\cite{7496808} and TaO\textsubscript{x}-HfO\textsubscript{x}~\cite{8510690}) to match or exceed the performance of high-accuracy benchmarks such as EpiRAM~\cite{c30a52423ab14e6c83bcac6aa8cae784}. 
Specifically, for TaO\textsubscript{x}-HfO\textsubscript{x} we consistently observe at least 2 orders of magnitude improvement in energy use and latency. Our results demonstrate that MELISO+ enables reliable and scalable in-memory computation, advancing the practical viability of RRAM-based in-memory computing for a general class of high-dimensional and high-precision computation tasks.

\begin{figure}[!htbp]
    \centering
    \begin{minipage}[t]{\textwidth}
        \centering
        \subfigure{%
            \includegraphics[width=\textwidth]{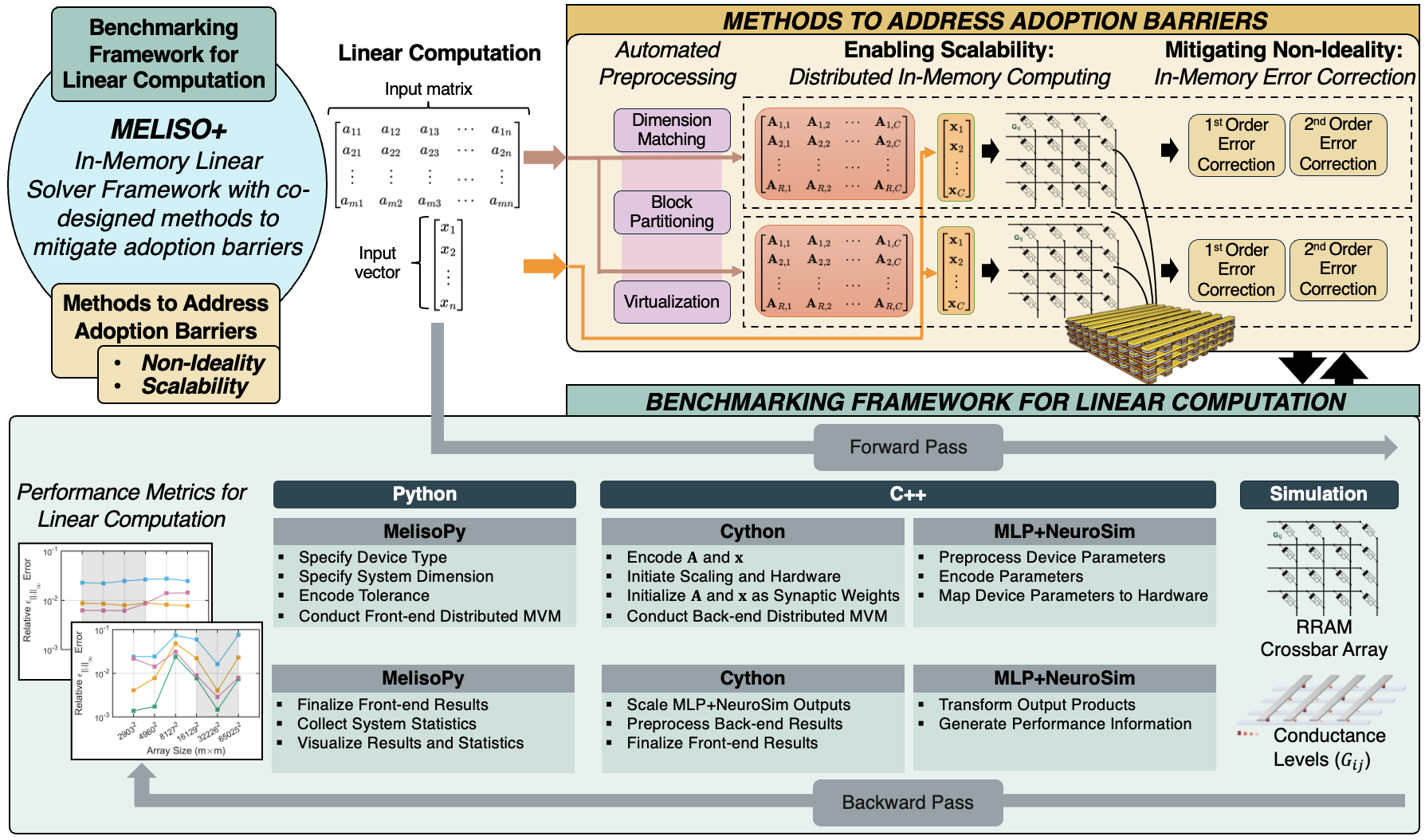}%
        }
    \end{minipage}
    \hfill
    \caption{Overview of the MELISO+~\cite{10766540} framework for RRAM computation with matrix-vector multiplication benchmark. 
    }
\label{fig:overview}
\end{figure}

\section{Results}\label{sec:Results}

We develop a benchmarking framework for scalable general-purpose linear computation tasks called MELISO+ (In-Memory Linear Solver), which builds on our previously reported platform~\cite{10766540}, to specifically address scalability and variability challenges. MELISO+ is a full-stack benchmarking, simulation and error correction framework designed for emulating analog matrix-vector multiplication (MVM) operations under realistic device conditions to evaluate a range of parameters related to variability, accuracy, and computational scalability. 

As illustrated in Fig.~\ref{fig:overview}, MELISO+ comprises a modular simulation pipeline with two computational stages: a forward pass for MVM and a backward pass for performance analysis and statistical evaluation. Inputs such as matrix dimensions, device type, and crossbar tolerance levels are specified using a Python interface (MelisoPy), which then communicates with a C++ engine built on top of NeuroSim+ \cite{8268337} to simulate MVM execution across RRAM arrays. The output is processed and scaled back into a Python environment, providing insights into generated MVM errors and high-level performance metrics. This flexible, end-to-end design supports hardware-software co-design and accelerates development cycles for energy-efficient linear solvers across a range of device technologies. 


\subsection{Linear computation, scalability and performance evaluation using RRAM devices}\label{sec:Results.1}

We focus on linear computation tasks as a primary application since they form the backbone of all modern computing. 
Without loss of generality, we consider a linear operation, denoted by $\mathbf{Ax} = \mathbf{b}$, where we compute the multiplication of a given matrix $\mathbf{A}$ with a given vector $\mathbf{x}$ and acquire the resulting vector $\mathbf{b}$. This computational task is a foundational operation that lies at the core of every modern computation task, ranging from artificial intelligence to constrained optimization. During computation, the matrix $\mathbf{A}$ and the vector $\mathbf{x}$ will be encoded to the RRAM, which will inevitably introduce some noise (i.e., inaccuracies). We denote their encoded counterparts as $\mathbf{\tilde{A}}$, and $\mathbf{\tilde{x}}$, respectively. The matrix $\mathbf{A}$ and vector $\mathbf{x}$ are incrementally programmed onto the RRAM crossbar using a closed-loop~\texttt{adjustableWriteandVerify} protocol~\cite{8894568}. The RRAM implementation of the method iteratively perturbs conductance values until the encoded representations, denoted $\tilde{\mathbf{A}}$ and $\tilde{\mathbf{x}}$, fall within a specified tolerance range or until a maximum iteration count is reached. 

Our proposed framework handles error correction and computational scalability using two separate but complementary methods. The error correction method mitigates first- and second-order errors to improve accuracy, while the scalability method builds on an adoption of a parallel computing framework for RRAMs. Together, these methods enable scalable and reliable computing, two significant roadblocks for adoption of RRAM technology.

The proposed error correction strategy builds on a two-stage algorithmic scheme designed to mitigate both first- and second-order arithmetic inaccuracies. In the first stage, the proposed method employs the structural properties of the linear operation to mitigate the first-order error (further discussed in the \nameref{sec:Methods} section). During the implementation, three matrix-vector products are computed: $\mathbf{A}\tilde{\mathbf{x}}$, $\tilde{\mathbf{A}}\mathbf{x}$, and $\tilde{\mathbf{A}}\tilde{\mathbf{x}}$. The circuit-level implementation proceeds as follows:
\begin{itemize}
    \item We construct a matrix $\mathbf{X}^\top \in \mathbb{R}^{n \times n}$ where each row is equal to $\mathbf{x}^\top$, i.e., $ \mathbf{X}^\top = [\mathbf{x}^\top | \mathbf{x}^\top | \dots | \mathbf{x}^\top]$ with $n$ identical rows. This matrix is written to the RRAM array. The transpose of each column of $\mathbf{A}$ (i.e., $\mathbf{a}_j^\top$ where $\mathbf{a}_j$ is a column of $\mathbf{A}$) is then provided as an input vector and multiplied with the corresponding row of $\mathbf{X}^\top$ to produce the output $ \mathbf{A}\tilde{\mathbf{x}}$. The vector $\tilde{\mathbf{x}}$ is noisy due to errors introduced during the encoding of $\mathbf{x}$. The resulting entries of $\tilde{\mathbf{X}}^\top$ are read and recorded.
    \item The matrix $\mathbf{A}$ is then encoded in the same RRAM array, replacing the previous entries of $\tilde{\mathbf{X}}^\top$. The exact vector $\mathbf{x}$ is provided as an input vector to compute $\tilde{\mathbf{A}}\mathbf{x}$. 
    \item Following up, the previously recorded $\mathbf{X}^\top$ values are retrieved row-wise and used as inputs to compute the product with the corresponding columns of $\tilde{\mathbf{A}}$, yielding $\tilde{\mathbf{A}}\tilde{\mathbf{x}}$. 
    \item Finally, these terms are combined to suppress first-order errors as:
    
    \begin{equation*}
        \mathbf{p} = \tilde{\mathbf{A}}\mathbf{x} + \mathbf{A}\tilde{\mathbf{x}} - \tilde{\mathbf{A}}\tilde{\mathbf{x}}.
    \end{equation*}

\end{itemize}

The proposed reformulation effectively cancels the dominant first-order error terms introduced during the programming process. In the second stage, the intermediate result $\mathbf{p}$ is further improved using a regularized least-squares denoising algorithm (discussed in the \nameref{sec:Methods} section) that attenuates remaining second-order residuals while preserving the structure of the true solution. Specifically, we first encode the matrix $(\mathbf{I}_n + \lambda \mathbf{L}^\top \mathbf{L})^{-1}$, and use this encoded matrix to evaluate $\mathbf{y} (\lambda) = (\mathbf{I}_n + \lambda \mathbf{L}^\top \mathbf{L})^{-1} \mathbf{p}$
where: $\lambda \in (0,1)$ is a regularization parameter—we selected $\lambda = 10^{-12}$ in MELISO+ since it produced the best result; $\mathbf{L}$ is the first-order differential matrix; and $\mathbf{I}_n$ is an identity matrix of size $n\times n$.

This software-based correction strategy enables high-accuracy computation using noisy analog devices and generalizes across device types, error models, and computational workloads. 

To support large-scale computation on fixed-size RRAM crossbars, we implement a distributed computing backend that partitions:
the input matrix $\mathbf{A} \in \mathbb{R}^{m \times n}$ row-wise into submatrices, denoted by $\mathbf{A}^{(i)} \in \mathbb{R}^{m_i \times n_i}$ with $m_i < m$ and $n_i < n$;
the input vector $\mathbf{x} \in \mathbb{R}^{n \times 1}$ row-wise into subvectors, denoted by $\mathbf{x}^{(i)} \in \mathbb{R}^{n_i \times 1}$. Each pair of submatrix-subvector is then processed independently on a separate compute node using the Message Passing Interface (MPI) protocol. Each node performs its local MVM operation—that is, $\mathbf{y}^{(i)} = \mathbf{A}^{(i)}\mathbf{x}^{(i)}$, applies on-node error correction and returns the partial result for aggregation. To overcome hardware constraints on array size and conductance range, we incorporate a virtualization layer that encodes arbitrarily large matrices onto fixed crossbar configurations by serializing sub-blocks and managing address mappings in software. This approach enables full-system simulation of matrix dimensions exceeding $65{,}000 \times 65{,}000$, with linear scaling in both memory and runtime across nodes. Together, distributed execution and virtualization allow MELISO+ to evaluate RRAM-based in-memory computing at a scale relevant for real-world scientific and energy applications.

We conducted a series of experiments to showcase the computational accuracy and scalability of the proposed method. We evaluated the computation accuracy performance of these devices using four key metrics: relative $\ell_2$-norm and $\ell_\infty$-norm errors, denoted by $\pmb{\epsilon}_{||.||_2}$ and $\pmb{\epsilon}_{||.||_\infty}$, respectively; write energy, denoted by $\text{E}_{\text{w}}$ and measured in joules (J); finally, write latency, denoted by $\text{L}_{\text{w}}$ and measured in seconds (s).
These relative error norms are defined as follows:

\begin{equation*} 
\pmb{\epsilon}_{\text{total}} = \frac{||\mathbf{y} - \mathbf{b}||_{p}}{||\mathbf{b}||_{p}}, \quad p \in \{2, \infty\} 
\end{equation*}

where $\mathbf{y}$ is the in-memory final result, $\mathbf{b}$ is the ground-truth result of the linear operation, $\mathbf{A}\mathbf{x} = \mathbf{b}$, and $p$ is the user-specific norm ($\ell_2$ or $\ell_\infty$).


\subsection{Benchmarking the error correction method}\label{subsection:4.2}
Here, we assess the impact of our proposed  first- and second-order error correction method on the MCAs. These devices were fabricated from four distinct material systems: Ag-aSi~\cite{doi:10.1021/nl904092h}, AlO\textsubscript{x}-HfO\textsubscript{2}~\cite{7496808}, EpiRAM~\cite{c30a52423ab14e6c83bcac6aa8cae784}, and TaO\textsubscript{x}-HfO\textsubscript{x}~\cite{8510690}. As the performance of the~\texttt{adjustableWriteandVerify} scheme depends on the number of iterations, we conducted experiments with iteration counts ranging from $k = 0$ to $k = 20$. For each iteration count, we repeated the experiments 100 times to ensure statistical reliability. 

These experiments utilized two matrices, \textbf{bcsstk02} (obtained from the SuiteSparse Matrix Collection~\cite{Kolodziej2019}) and \textbf{Iperturb} (a slightly perturbed identity matrix), which are of comparable size (both are square matrices of size $66\times66$) but exhibit distinct condition numbers, denoted by $\kappa(\mathbf{A})$, enabling us to evaluate the methods under varying levels of numerical stability. The input vector $\mathbf{x}\in\mathbb{R}^{66\times1}$ employed in these experiments is constructed such that it is a single observation sampled from the standard multivariate normal distribution—that is, $\mathbf{x} \sim \mathcal{N}(\mathbf{0},\sigma^2\mathbf{I}_{66})$ with $\sigma^2 =1$.

Table~\ref{tabular:errorCorrection&setWeightsIncremental} compares the performance of Ag-aSi, AlO\textsubscript{x}-HfO\textsubscript{2}, and TaO\textsubscript{x}-HfO\textsubscript{x} devices when subjected to our proposed error correction methods and the~\texttt{adjustableWriteandVerify} scheme. EpiRAM devices that innately have significantly higher accuracy serve as the benchmark for regular device performance without error correction. Table~\ref{tabular:errorCorrection&setWeightsIncremental} presents results containing averaged metrics over 100 experimental replications. Results from Table~\ref{tabular:errorCorrection&setWeightsIncremental} indicate that the proposed error correction methods significantly reduce relative error norms in MVM across all tested devices for matrices with varying condition numbers (i.e. M1 and M2), thereby mitigating error propagation. 
Although error correction implementation increases both writing energy, $\text{E}{\text{w}}$, and writing latency, $\text{L}{\text{w}}$ compared to a direct computation, it provides two distinct advantages: (i) with error correction, error terms achieved by lower accuracy devices suchs as TaO$_x$-HfO$_x$ become on par with the de facto best-performing device, EpiRAM, and (ii) the resulting TaO$_x$-HfO$_x$ computation still provides approximately 3–5 orders of magnitude reduction in energy and 2 orders of magnitude reduction in latency in comparison to an EpiRAM computation. 
This demonstrates that devices with low latency and energy consumption—but higher intrinsic error rates due to lower synaptic weight precision or variability, e.g., TaO$_x$-HfO$_x$—can achieve comparable accuracy to the best-performing RRAM devices when paired with the proposed error correction framework, while offering significant advantages in both energy and latency.

\begin{table}[h] 
\centering
\captionsetup{labelfont=bf,
  textfont=bf,justification=raggedright,
  singlelinecheck=false}
\caption{Device performances for MVM operations with and without the proposed error correction method, using EpiRAM as the benchmark.}
\medskip
\small
\begin{tabular}{p{0.5cm}cc|ccc|ccc|}
\multicolumn{1}{c}{}        & \rot{\textbf{Metrics}}    
& \rot{\textbf{EpiRAM}}   & \rot{\textbf{Ag-aSi}}           & \rot{\textbf{AlO$_x$-HfO$_2$}}       & \rot{\textbf{TaO$_x$-HfO$_x$}}     & \rot{\textbf{Ag-aSi}}        & \rot{\textbf{AlO$_x$-HfO$_2$}}  & \rot{\textbf{TaO$_x$-HfO$_x$}}   
\\
\hline
\multicolumn{1}{c|}{\multirow{4}{*}{\rotatebox{90}{\textbf{M1}}}} 
&\multicolumn{1}{c|}{$\pmb{\epsilon}_{||.||_2}$}     & 0.0223  & 0.2305          & 0.6001          & 0.4914          & 0.0350        & 0.0204         & 0.0300         \\ 
\multicolumn{1}{c|}{} &
\multicolumn{1}{c|}{$\pmb{\epsilon}_{||.||_\infty}$}     & 0.0261  & 0.2327          & 0.5991          & 0.4787          & 0.0417        & 0.0298         & 0.0321         \\ 
\multicolumn{1}{c|}{} &
\multicolumn{1}{c|}{$\text{E}_{\text{w}}$}     & 0.0001  & 3.75E-06        & 5.52E-05        & 5.36E-08        & 5.12E-06      & 0.0001        & 7.48E-08       \\
\multicolumn{1}{c|}{} &
\multicolumn{1}{c|}{$\text{L}_{\text{w}}$}     & 0.0449  & 1.0089          & 0.1398          & 0.0002          & 1.3652        & 0.2227         & 0.0003         \\ \hline
\multicolumn{1}{c|}{\multirow{4}{*}{\rotatebox{90}{\textbf{M2}}}} & 
\multicolumn{1}{c|}{$\pmb{\epsilon}_{||.||_2}$}   & 0.6383  & 1.1364          & 1.0546          & 1.1468          & 0.1758        & 0.2064         & 0.4428         \\
 \multicolumn{1}{c|}{}   &
\multicolumn{1}{c|}{$\pmb{\epsilon}_{||.||_\infty}$}     & 0.5466  & 1.0389          & 1.0155          & 1.0416          & 0.1424        & 0.1628         & 0.3035         \\
 \multicolumn{1}{c|}{}   &
\multicolumn{1}{c|}{$\text{E}_{\text{w}}$}     & 1.19E-05 & 3.04E-07        & 4.83E-06        & 4.71E-09        & 3.70E-07      & 7.69E-05       & 2.31E-08       \\
 \multicolumn{1}{c|}{}   &
\multicolumn{1}{c|}{$\text{L}_{\text{w}}$}     & 0.0428  & 0.9507          & 0.1320          & 0.0002          & 1.8884        & 0.2772         & 0.00028        \\ \hline

\hline
\multicolumn{3}{l|}{} & \multicolumn{3}{c|}{\textbf{No Error Correction}} & \multicolumn{3}{c|}{\textbf{With Error Correction}} 
\\
\cmidrule{4-9}

\multicolumn{9}{l}{
  \begin{tabular}[l]{@{}l@{}}
    $\pmb{\epsilon}_{||.||_2}$ and $\pmb{\epsilon}_{||.||_\infty}$ are relative $\ell_2$- and $\ell_\infty$-norm error. $\text{E}_{\text{w}}$ and $\text{L}_{\text{w}}$ are write energy, and latency.\\
    M1 is the \textbf{bcsstk02} matrix with $\kappa(\mathbf{A}) = 4324.91$, and $\text{dim}(\mathbf{A})=66\times66$
    \\
    M2 is the \textbf{Iperturb} matrix with $\kappa(\mathbf{A})=1.2342$, and $\text{dim}(\mathbf{A})=66\times66$
    \\
    All devices (excluding EpiRAM) are subjected to first- and second-order error correction.\\
    All devices (including EpiRAM) are subjected to the~\texttt{adjustableWriteandVerify} scheme.
  \end{tabular}
}
\end{tabular}
\label{tabular:errorCorrection&setWeightsIncremental}
\end{table}

\begin{figure*}[h]
    \centering
    \begin{minipage}[t]{0.49\textwidth}
        \centering
        \subfigure[]{%
            \includegraphics[width=\textwidth]{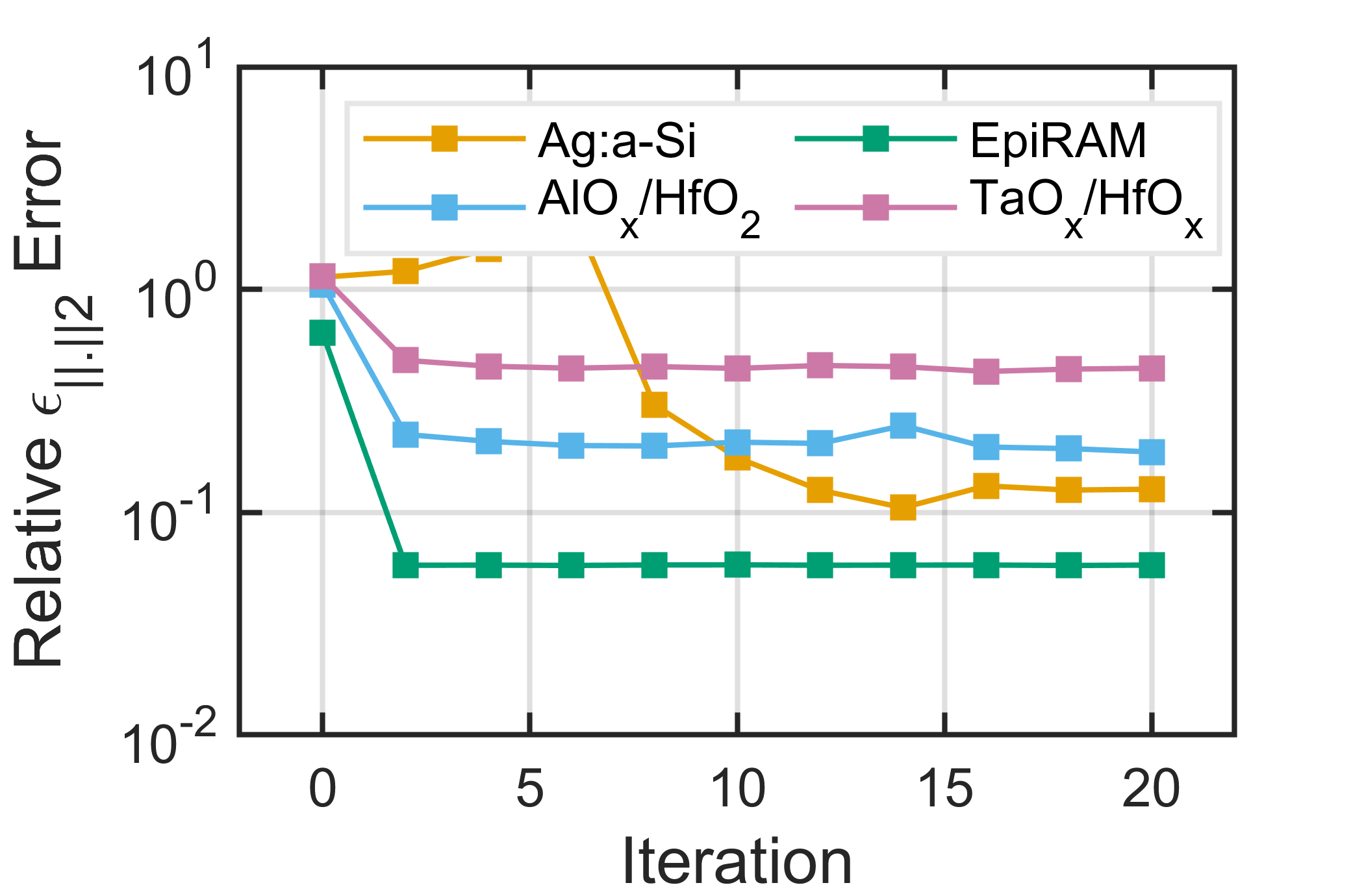}%
        }
    \end{minipage}
    \hfill
    \begin{minipage}[t]{0.49\textwidth}
        \centering
        \subfigure[]{%
            \includegraphics[width=\textwidth]{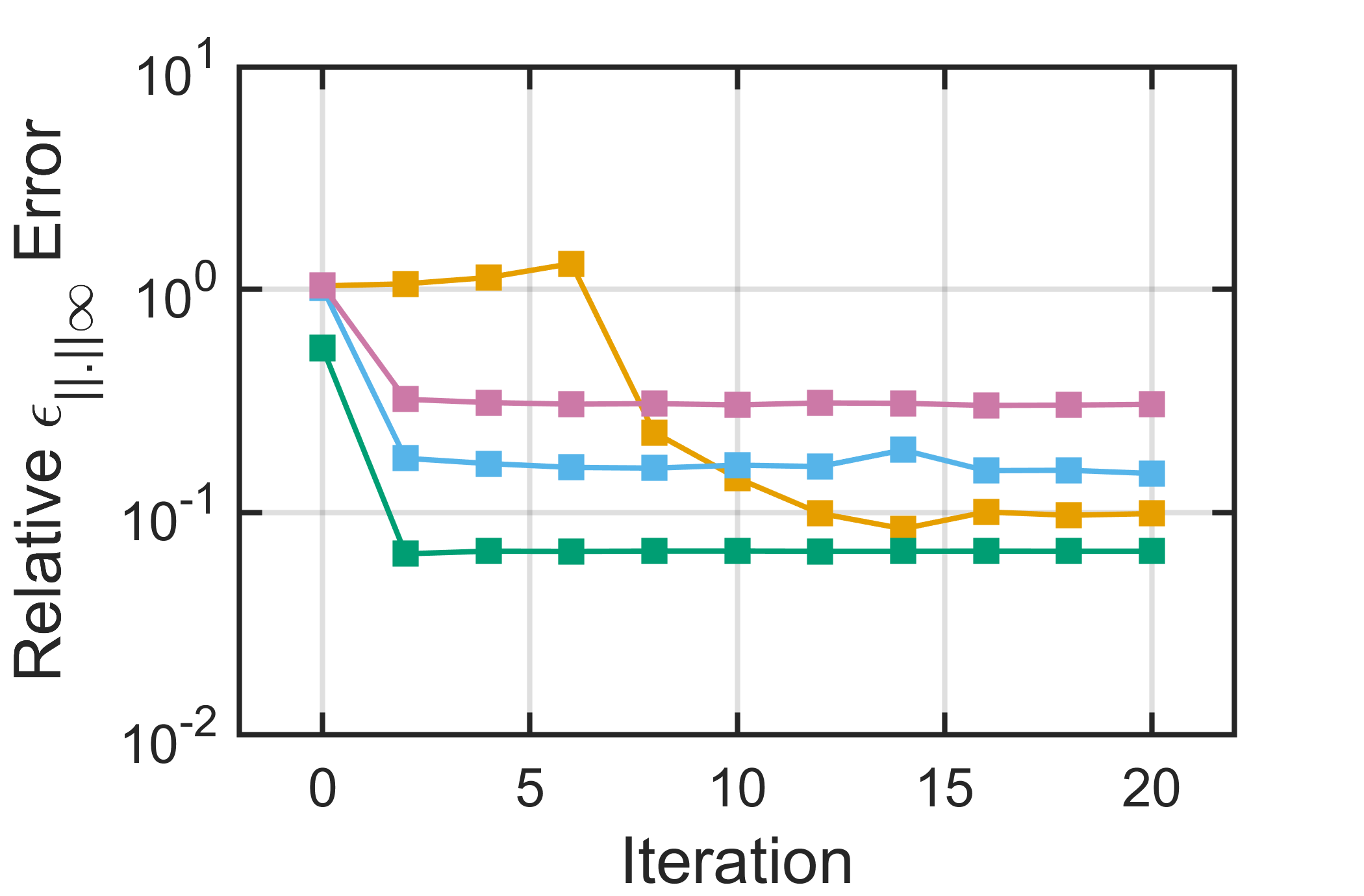}%
        }
    \end{minipage}
    
    \begin{minipage}[t]{0.49\textwidth}
        \centering
        \subfigure[]{%
            \includegraphics[width=\textwidth]{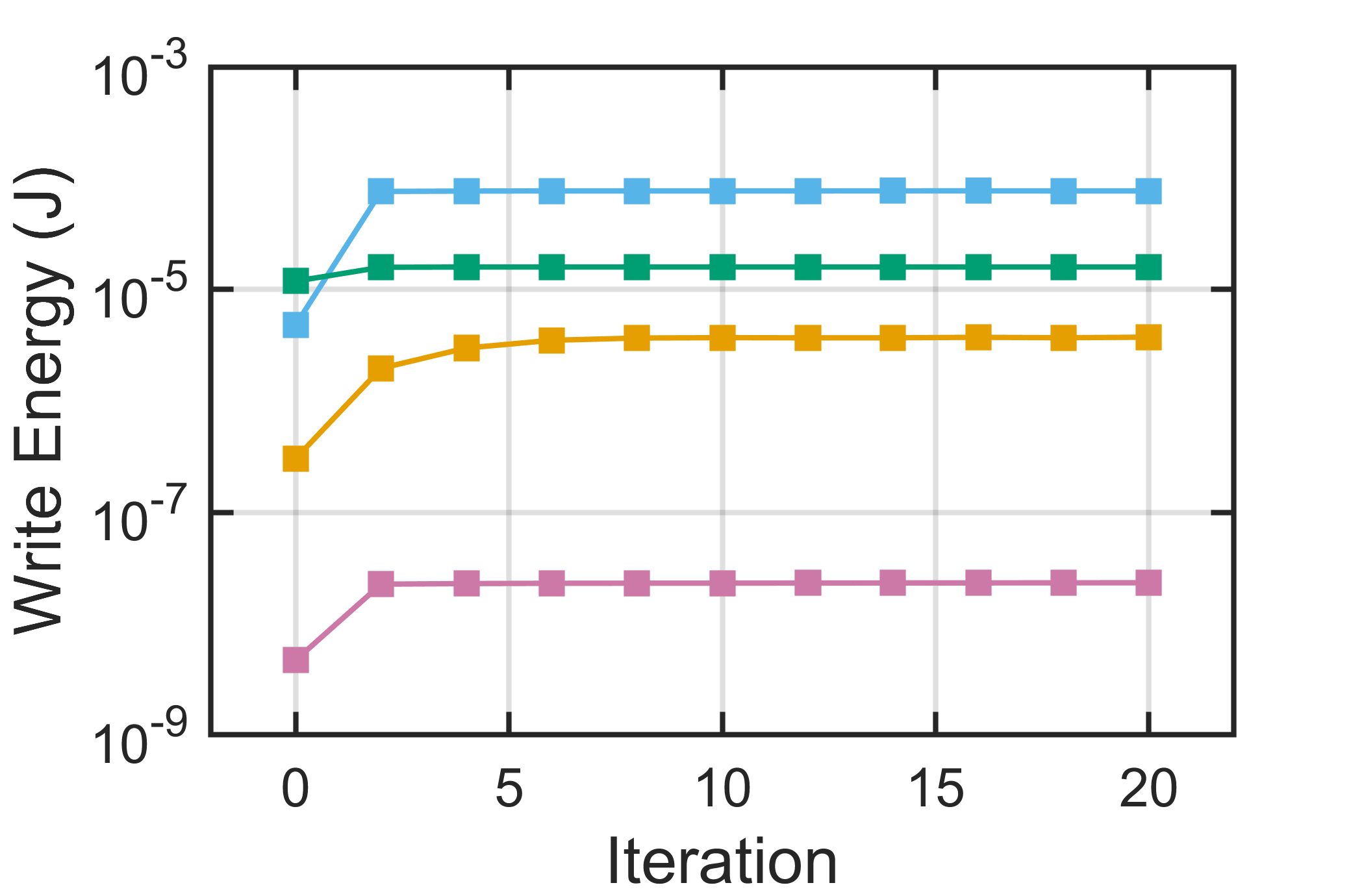}%
        }
    \end{minipage}
    \hfill
    \begin{minipage}[t]{0.49\textwidth}
        \centering
        \subfigure[]{%
            \includegraphics[width=\textwidth]{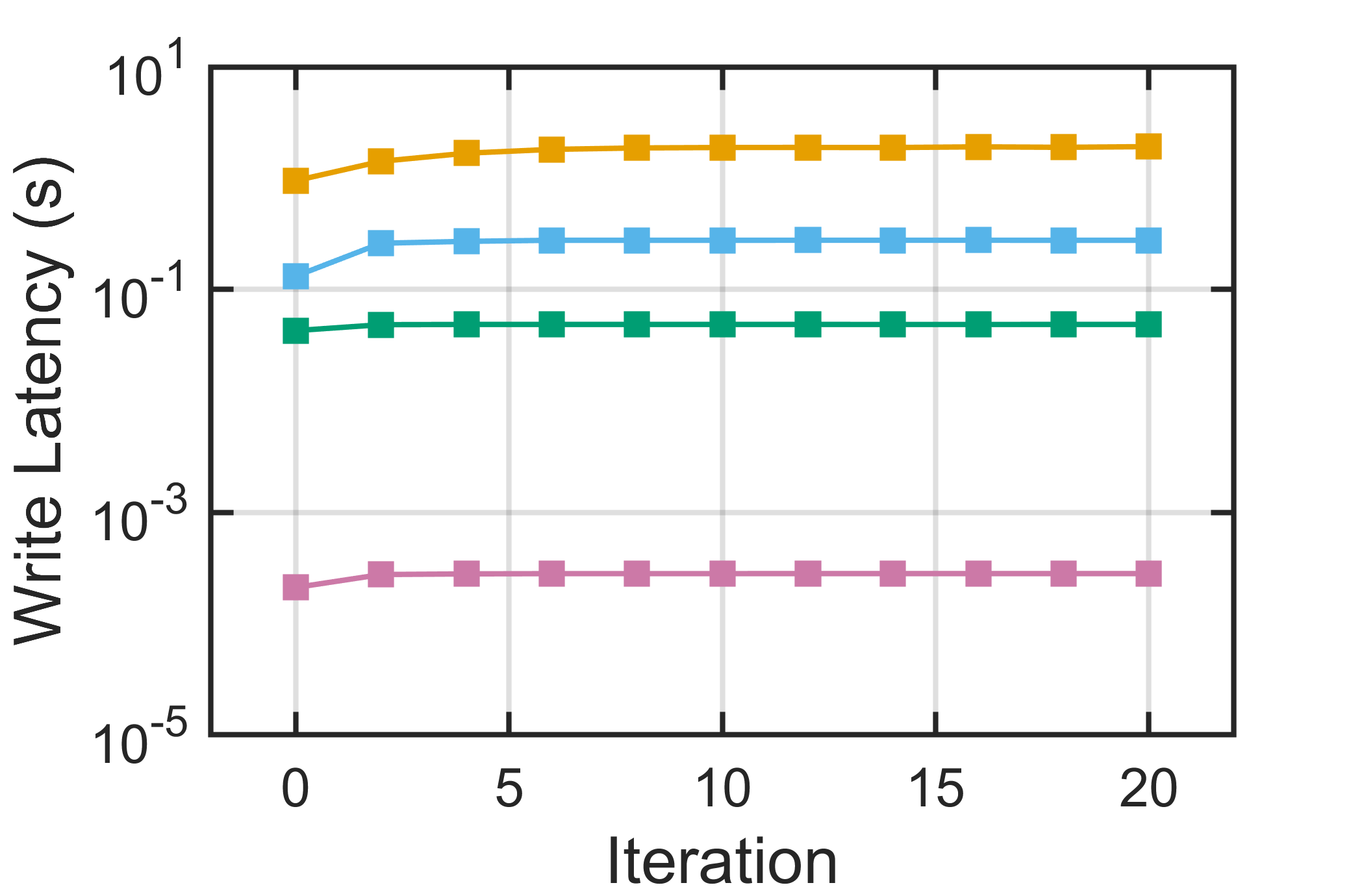}%
        }
    \end{minipage}
    
    \caption{Effects of the~\texttt{adjustableWriteandVerify} function (with no error correction algorithms applied) on the \textbf{Iperturb} matrix with different fixed numbers of iteration counts: (a) Relative $\ell_2$-norm Error, (b) Relative $\ell_\infty$-norm Error, (c) Write Energy, (d) Write Latency.}
    \label{fig:Iperturb_EC_0}
\end{figure*}

\begin{figure*}[h]
\centering
    \begin{minipage}[t]{0.49\textwidth}
        \centering
        \subfigure[]{%
            \includegraphics[width=\textwidth]{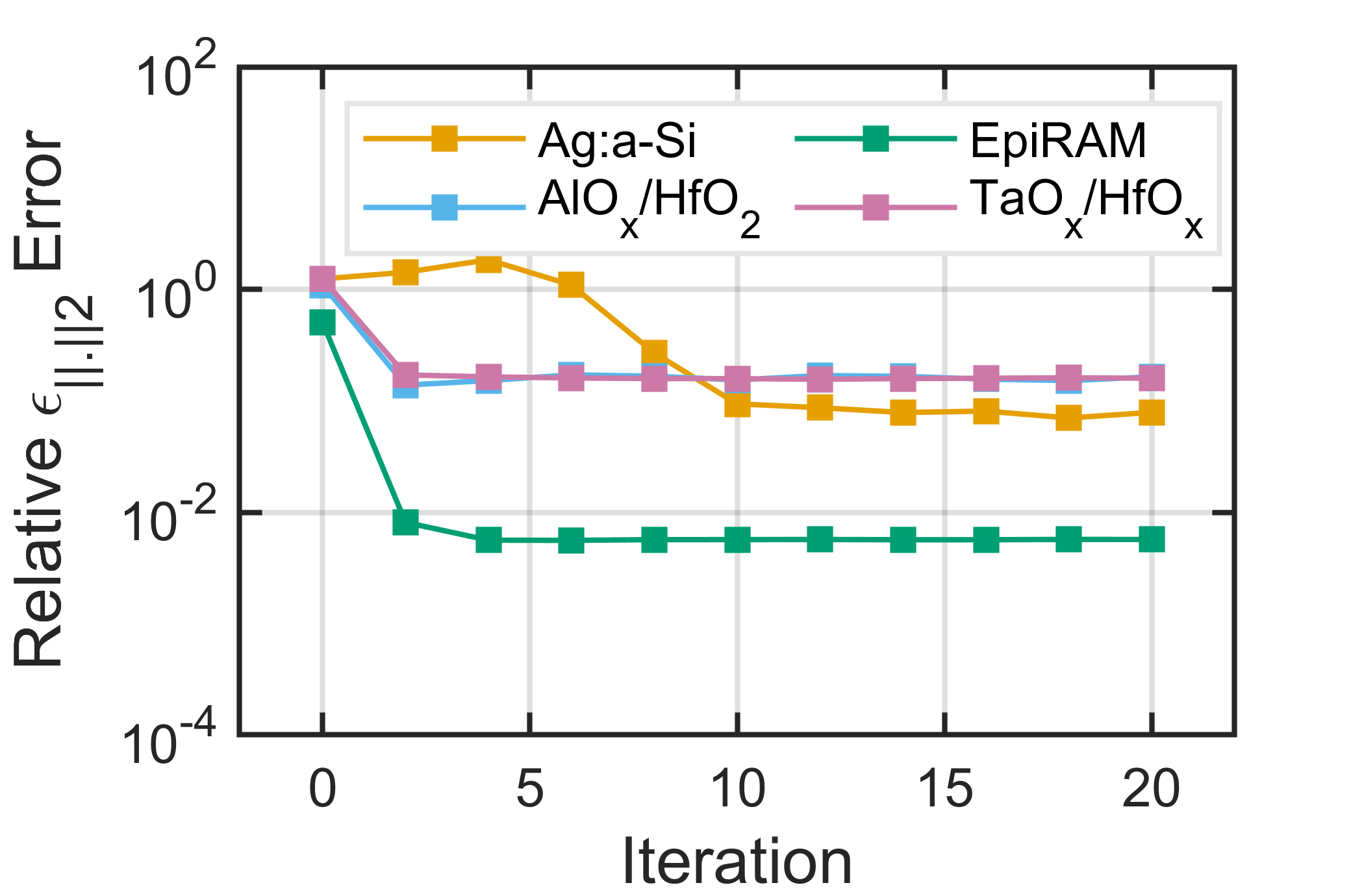}%
        }
    \end{minipage}
    \hfill
    \begin{minipage}[t]{0.49\textwidth}
        \centering
        \subfigure[]{%
            \includegraphics[width=\textwidth]{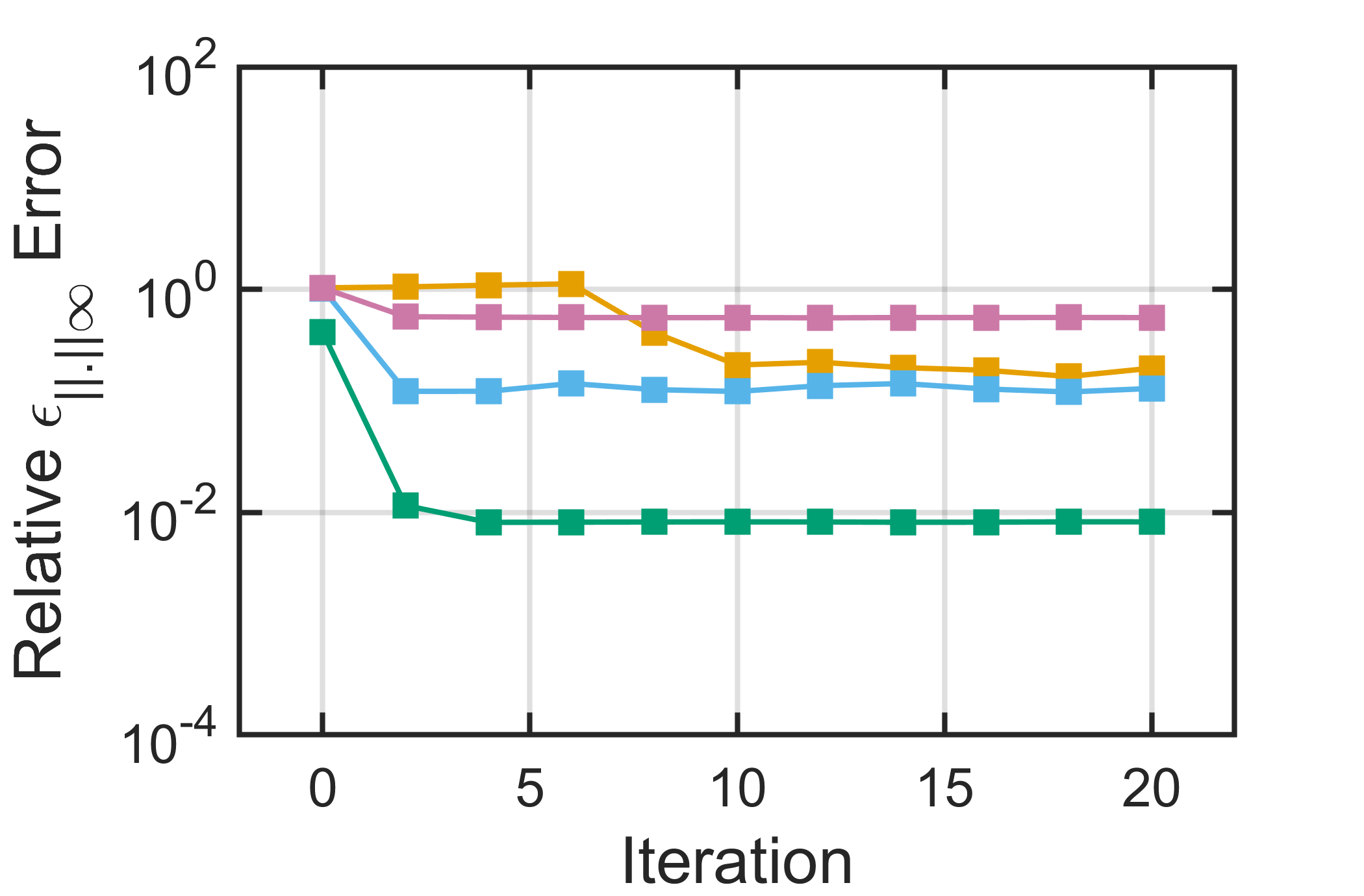}
        }
    \end{minipage}
    
    \begin{minipage}[t]{0.49\textwidth}
        \centering
        \subfigure[]{%
            \includegraphics[width=\textwidth]{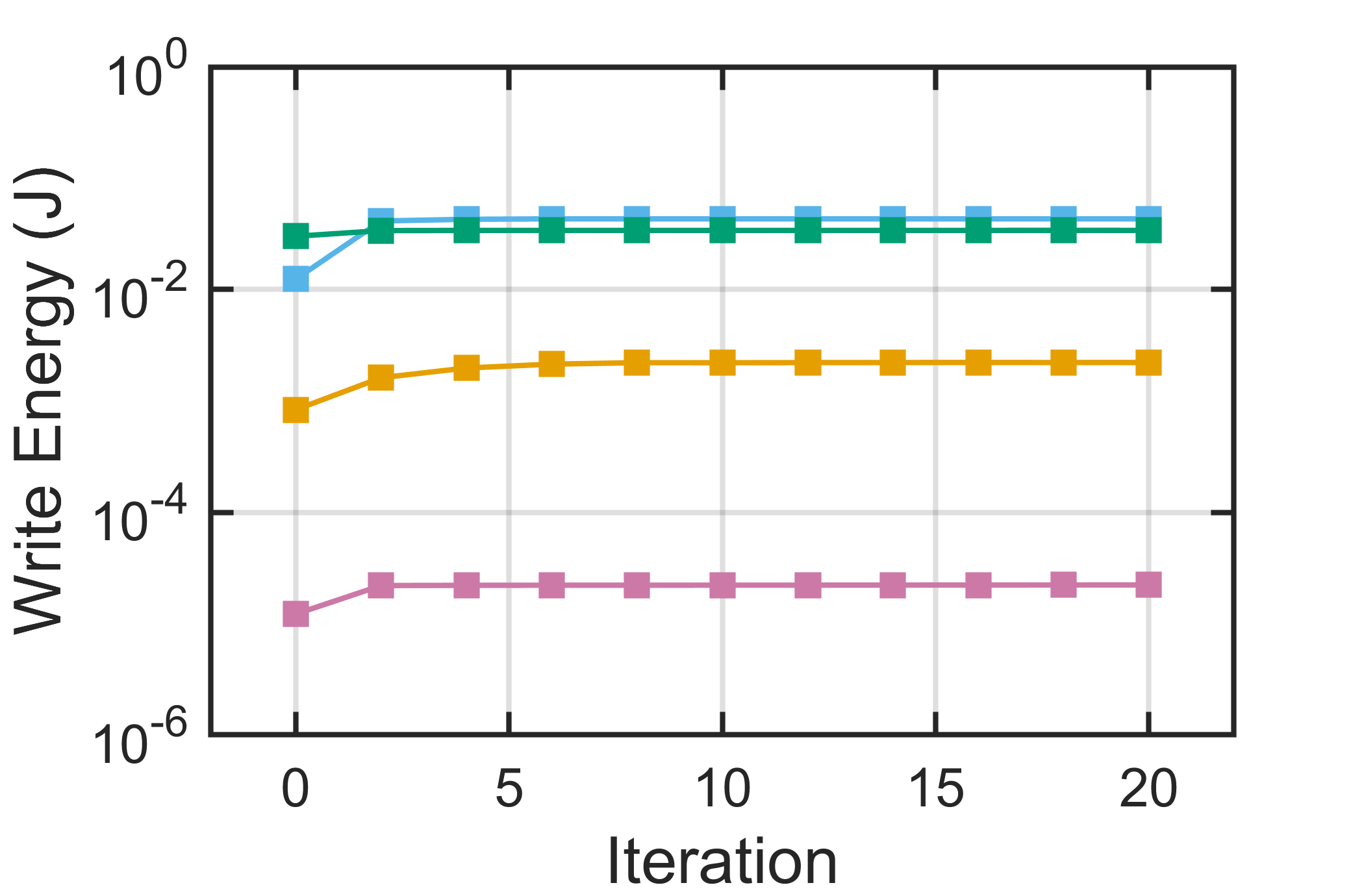}%
        }
    \end{minipage}
    \hfill
    \begin{minipage}[t]{0.49\textwidth}
        \centering
        \subfigure[]{%
            \includegraphics[width=\textwidth]{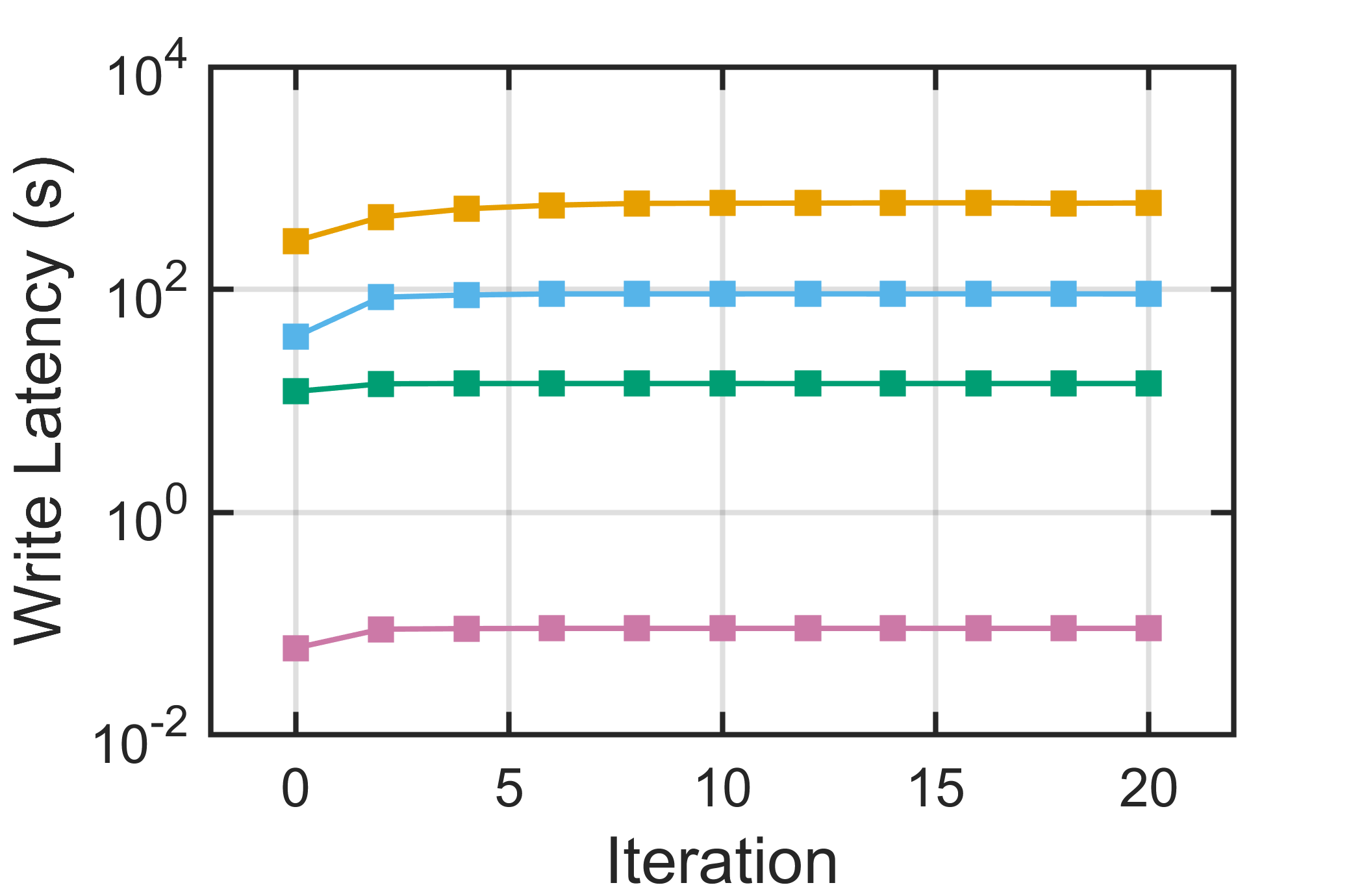}%
        }
    \end{minipage}
    \caption{Effects of the~\texttt{adjustableWriteandVerify} function (with error correction algorithms applied) on the \textbf{Iperturb} matrix with different fixed numbers of iteration counts: (a) Relative $\ell_2$-norm Error, (b) Relative $\ell_\infty$-norm Error, (c) Write Energy, (d) Write Latency.}
    \label{fig:Iperturb_EC_1}
\end{figure*}

Complementing the results from Table~\ref{tabular:errorCorrection&setWeightsIncremental}, Figs.~\ref{fig:Iperturb_EC_0} and~\ref{fig:Iperturb_EC_1} illustrate trends in these performance metrics, acquired from the MVM with \textbf{Iperturb} for $k=20$ iterations, without and with implemented error correction, respectively. (results for the MVM with \textbf{bcsstk02} are discussed in Supplementary Information~\ref{secA2}). 

Fig.~\ref{fig:Iperturb_EC_0} shows that applying only the~\texttt{adjustableWriteandVerify} scheme with multiple iterations ($k > 0$) significantly reduces relative error norms. For TaO\textsubscript{x}-HfO\textsubscript{x}, AlO\textsubscript{x}-HfO\textsubscript{2}, and EpiRAM, the reduction in relative error norms stabilizes at $k = 2$, while for Ag-aSi~\cite{doi:10.1021/nl904092h}, it stabilizes at $k = 11$. The distinct trend observed for Ag-aSi is attributed to the pronounced non-linearity ($2.4/-4.88$) in its synaptic weight update characteristics. 
When applied to ill-conditioned matrices (as detailed in Supplementary Information~\ref{secA1}), the proposed \texttt{adjustableWriteandVerify} method yields up to a two-orders-of-magnitude reduction in relative error norms. This enhanced performance is attributed to the algorithm’s ability to mitigate amplified numerical errors that arise from the high sensitivity of ill-conditioned systems to perturbations during analog weight programming.

Comparing Fig.~\ref{fig:Iperturb_EC_0}, which utilizes only the \texttt{adjustableWriteandVerify} scheme without error correction, and Fig.~\ref{fig:Iperturb_EC_1}, which combines \texttt{adjustableWriteandVerify} with the proposed error correction framework, demonstrate a significant further reduction in error norms with error correction that would not be possible through a write and verify scheme alone. 
Error reduction is an order of magnitude due to the inherently low condition number of the \textbf{Iperturb} matrix, where native matrix-vector multiplication (MVM) errors are intrinsically smaller. In contrast, for ill-conditioned matrices (as discussed in Supplementary Information~\ref{secA1}), integrating the proposed error correction strategy with the multi-iteration \texttt{adjustableWriteandVerify} scheme yields a substantially greater reduction in relative error norms, highlighting the synergistic effect of combining iterative programming with error-aware correction for more demanding numerical conditions.

These results demonstrate that a low-latency device, despite having lower accuracy, can potentially outperform an inherently high-accuracy device when subjected to the joint multi-iteration~\texttt{adjustableWriteandVerify} and error correction schemes, thus reducing the latency and energy needed to perform MVM with high accuracy.

\subsection{Scalability for Large-scale Applications}\label{subsec4.3}

A significant aspect of the proposed framework is the capability to scale-up using parallel computing. Building on the established benchmarking methodology for error correction effectiveness, we conducted scalability experiments for MELISO under two distinct paradigms:

\begin{itemize}
    \item Weak Scaling: This approach assesses the performance of MCA systems simulated using MELISO+ with a fixed problem size, presented by dimensions of input matrices, while scaling hardware resources. Specifically, we maintained a constant tile size in a system with multiple MCAs and varied the array sizes, increasing the number of RRAMs per array (referred to as array cell size), to simulate resource expansion.
    \item Strong Scaling: This approach evaluates the capacity of a multi-MCA system simulated by MELISO+ to accommodate progressively larger matrices using a fixed hardware configuration, where the tile size and array cell size are kept constant.
\end{itemize}


\subsubsection{Weak Scaling}\label{subsubsection:Weak-Scaling}
We investigated the weak-scaling capability of MELISO+ by performing MVM operations with error correction on a fixed matrix, \textbf{add32} (also obtained from the SuiteSparse Matrix Collection~\cite{Kolodziej2019}), whose dimensions are $4960 \times 4960$. 
The MCA cell size was incrementally increased from $32\times32$ to $1024\times1024$, emulating the scaling of hardware resources. The experimental system, which we referred to as the multi-MCA system, consisted of 64 MCAs configured in an $8\times8$ tile array. 

\begin{figure*}
    \centering
        \begin{minipage}[t]{0.49\textwidth}
            \centering
            \subfigure[]{\includegraphics[width=\textwidth]{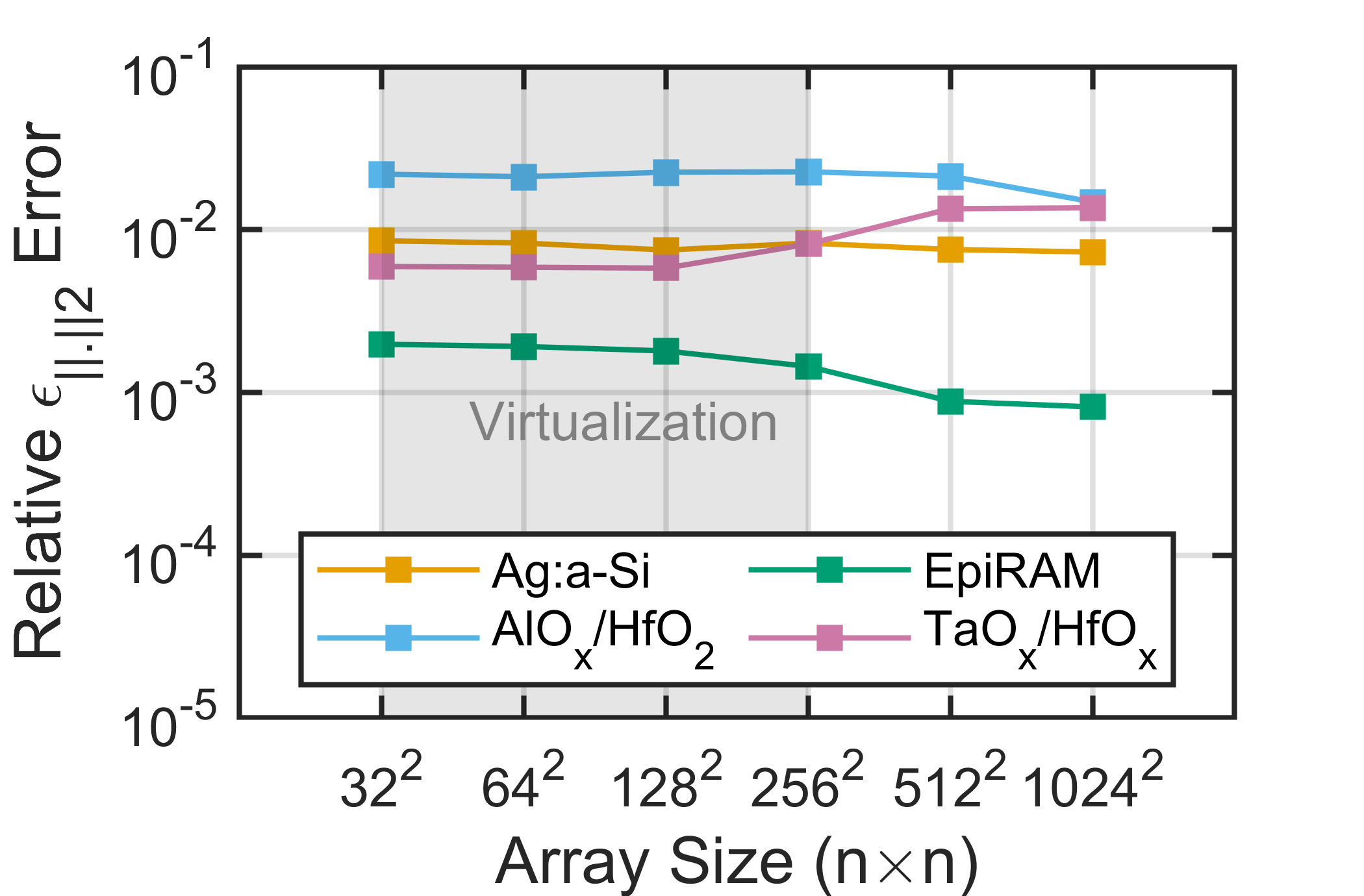}}
        \end{minipage}
        \hfill
        \begin{minipage}[t]{0.49\textwidth}
            \centering
            \subfigure[]{\includegraphics[width=\textwidth]{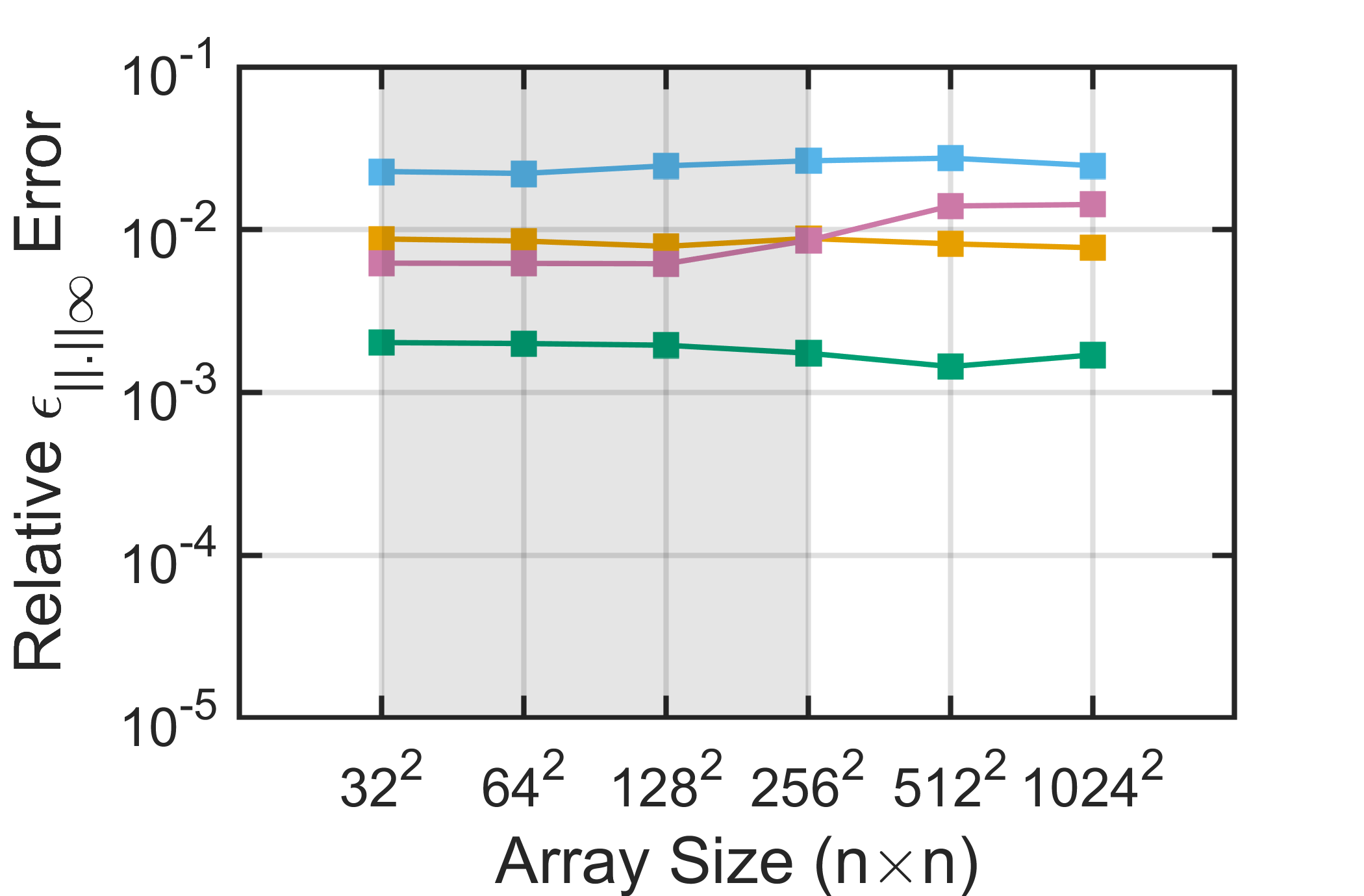}}
        \end{minipage}
    
        \begin{minipage}[t]{0.49\textwidth}
            \centering
            \subfigure[]{\includegraphics[width=\textwidth]{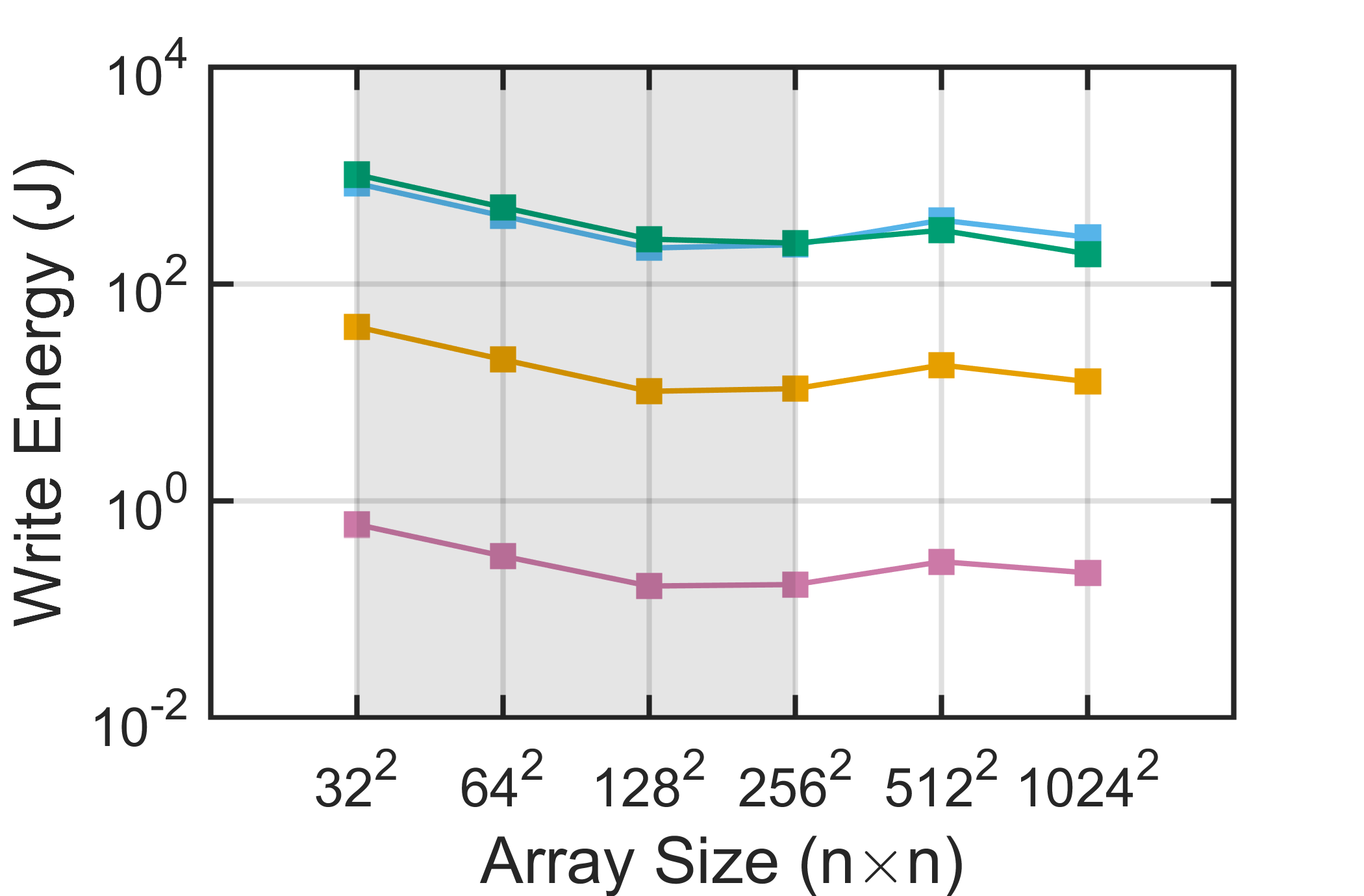}}
        \end{minipage}
        \hfill
        \begin{minipage}[t]{0.49\textwidth}
            \centering
            \subfigure[]{\includegraphics[width=\textwidth]{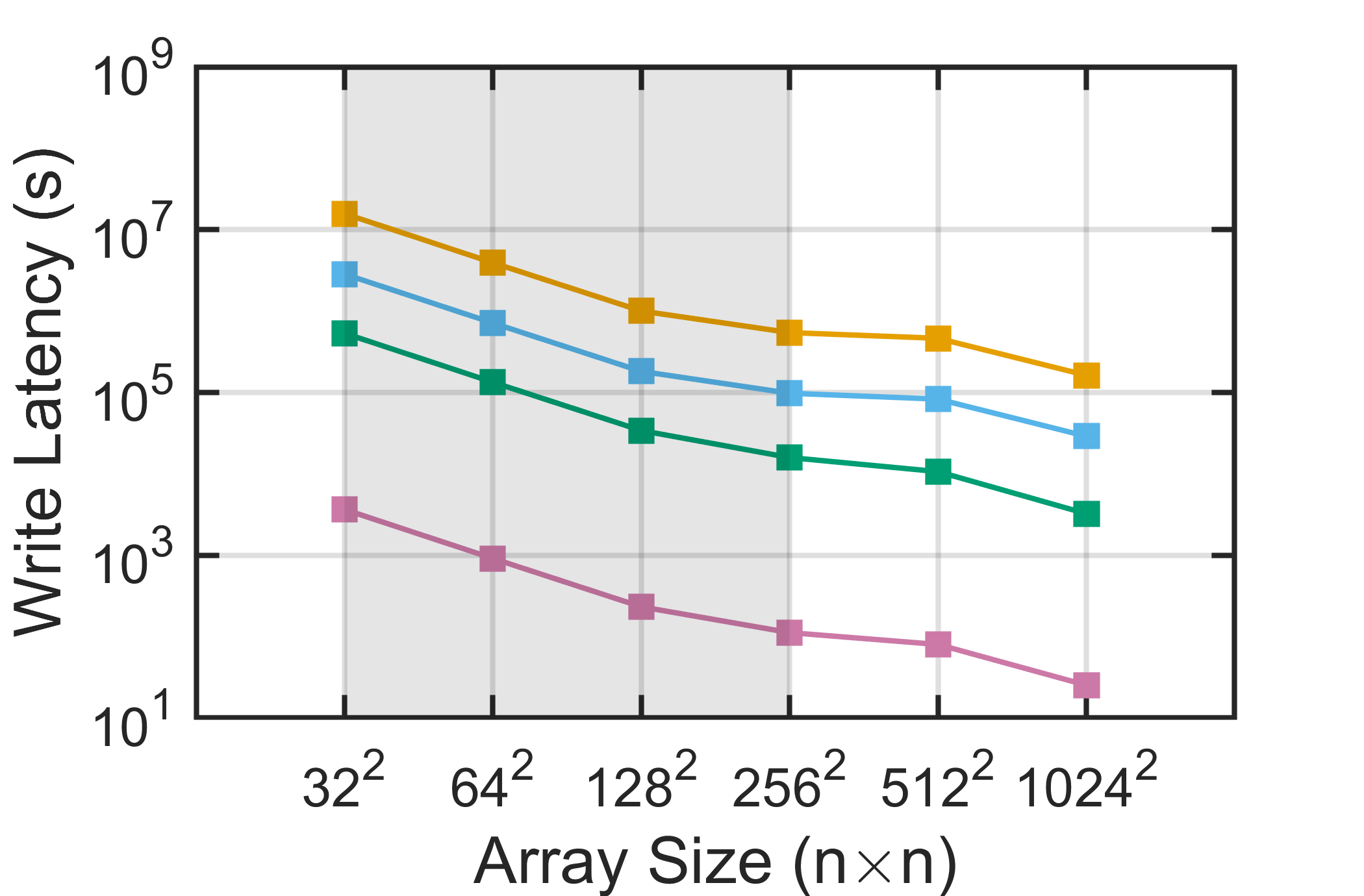}}
        \end{minipage}
    \caption{A comparison of device performance across key metrics: (a) Relative $\ell_2$-norm Error, (b) Relative $\ell_\infty$-norm Error, (c) Write Energy, (d) Write Latency. The performance is assessed on a fixed problem size using a multi-MCA system with a fixed tile size and varied different cell sizes per MCA. Due to the concurrent nature of distributing the MVM workload across a multi-MCA system of $8\times8$ tile sizes, write energy and write latency values were computed as the mean across all MCA devices.}\label{fig:commercialized_results} 
\end{figure*}

The performance of these systems, fabricated from the materials under study, is presented across varying array cell sizes in Fig.~\ref{fig:commercialized_results}. These results reveal that applying our proposed methods (multi-iteration~\texttt{adjustableWriteandVerify} together with first- and second-order error corrections) yields consistently low relative error norms, between $4*10^{-2}$ and $2*10^{-3}$ across all system configurations, underscoring the robustness and effectiveness of MELISO+ across diverse hardware setups. 
Fig.~\ref{fig:commercialized_results} also indicates a trend in the relationship between array cell size and system performance: smaller arrays (e.g., $32\times32$ to $256\times256$) exhibit significantly higher $\text{E}_{\text{w}}$ and $\text{L}_{\text{w}}$ compared to larger arrays (e.g., $512\times512$ to $1024\times1024$).

The performance disparity between smaller versus larger arrays stems from the virtualization paradigm implemented in MELISO+. When the physical dimensions of a multi-MCA system are smaller than the target matrix (due to reduced cell sizes), each MCA must be reassigned multiple times to complete the entire MVM operation. For example, in a multi-MCA system having an array cell size of $32\times32$ and a tile size of $8\times8$, each MCA is invoked 16 times, increasing computing overhead. In contrast, when the system’s physical dimensions match or exceed the problem size (e.g., with an array cell size of $512\times512$ or $1024\times1024$, while the tile size remains the same), virtualization is not applicable, enabling single-step execution with minimal error, energy consumption, and latency. These results underscore a critical insight: while our proposed methods consistently provide improvements in device performance, virtualization introduces trade-offs between hardware scalability versus computational efficiency. In other words, MELISO+ effectively preserves computational accuracy regardless of array size, with larger arrays offering superior energy and latency performance.

\subsubsection{Strong-Scaling}\label{subsubsection:Strong-Scaling}
In our strong scaling experiments, we assessed the capacity of MELISO+ to address larger MVM problems by computing the linear operation with different matrices of increasing dimensions obtained from the SuiteSparse Matrix Collection~\cite{Kolodziej2019} while maintaining a fixed hardware configuration. The details of each matrix is provided in Supplementary Information~\ref{secA1}.
For consistency with the methodology outlined in the previous section, we employed a multi-MCA system consisting of 64 MCAs arranged in a $8\times8$ tile, with each array configured to a constant array cell size of $1024\times1024$. 

\begin{figure*}
    \centering
        \begin{minipage}[t]{0.49\textwidth}
            \centering
            \subfigure[]{\includegraphics[width=\textwidth]{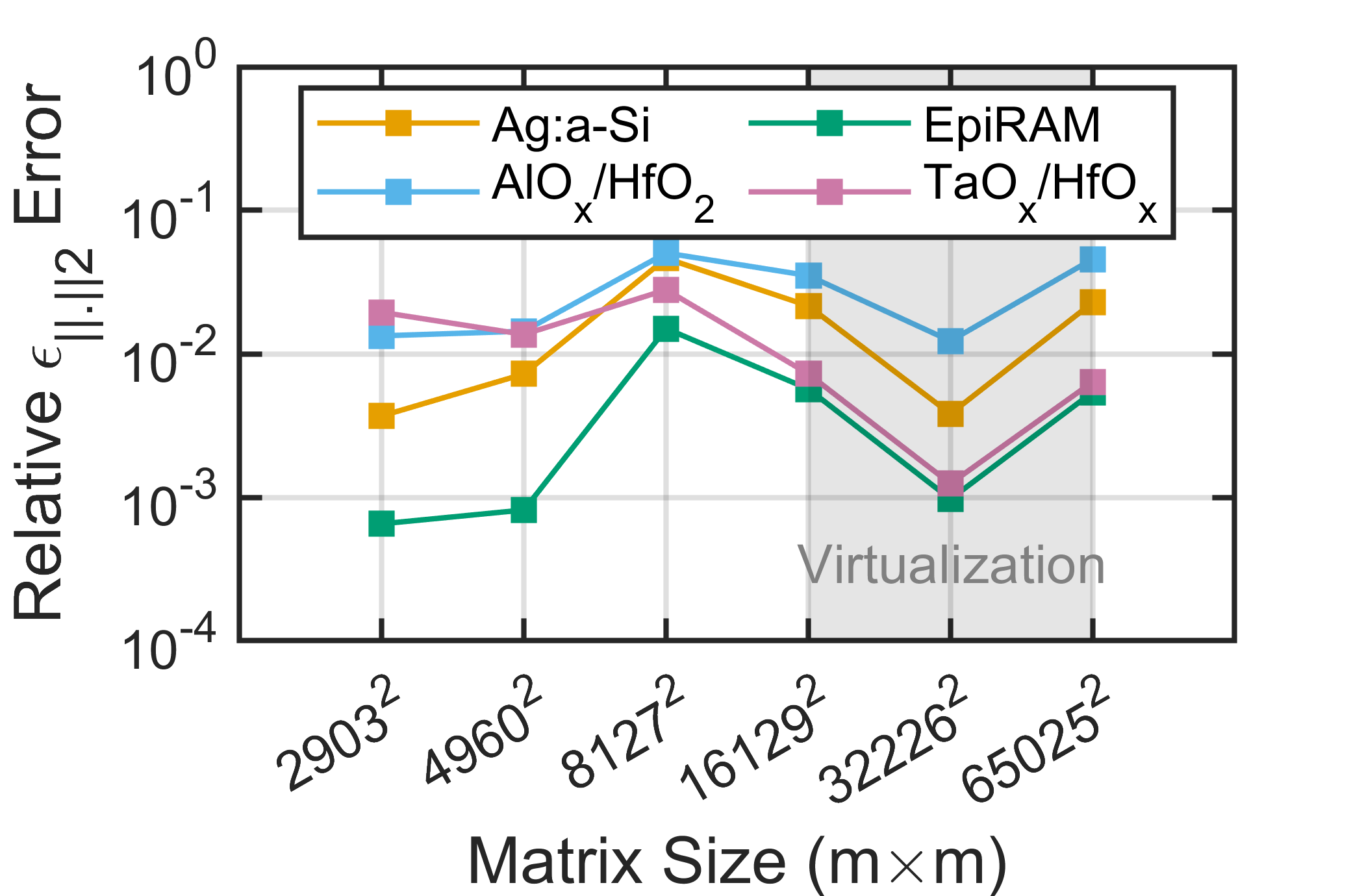}}
        \end{minipage}
        \hfill
        \begin{minipage}[t]{0.49\textwidth}
            \centering
            \subfigure[]{\includegraphics[width=\textwidth]{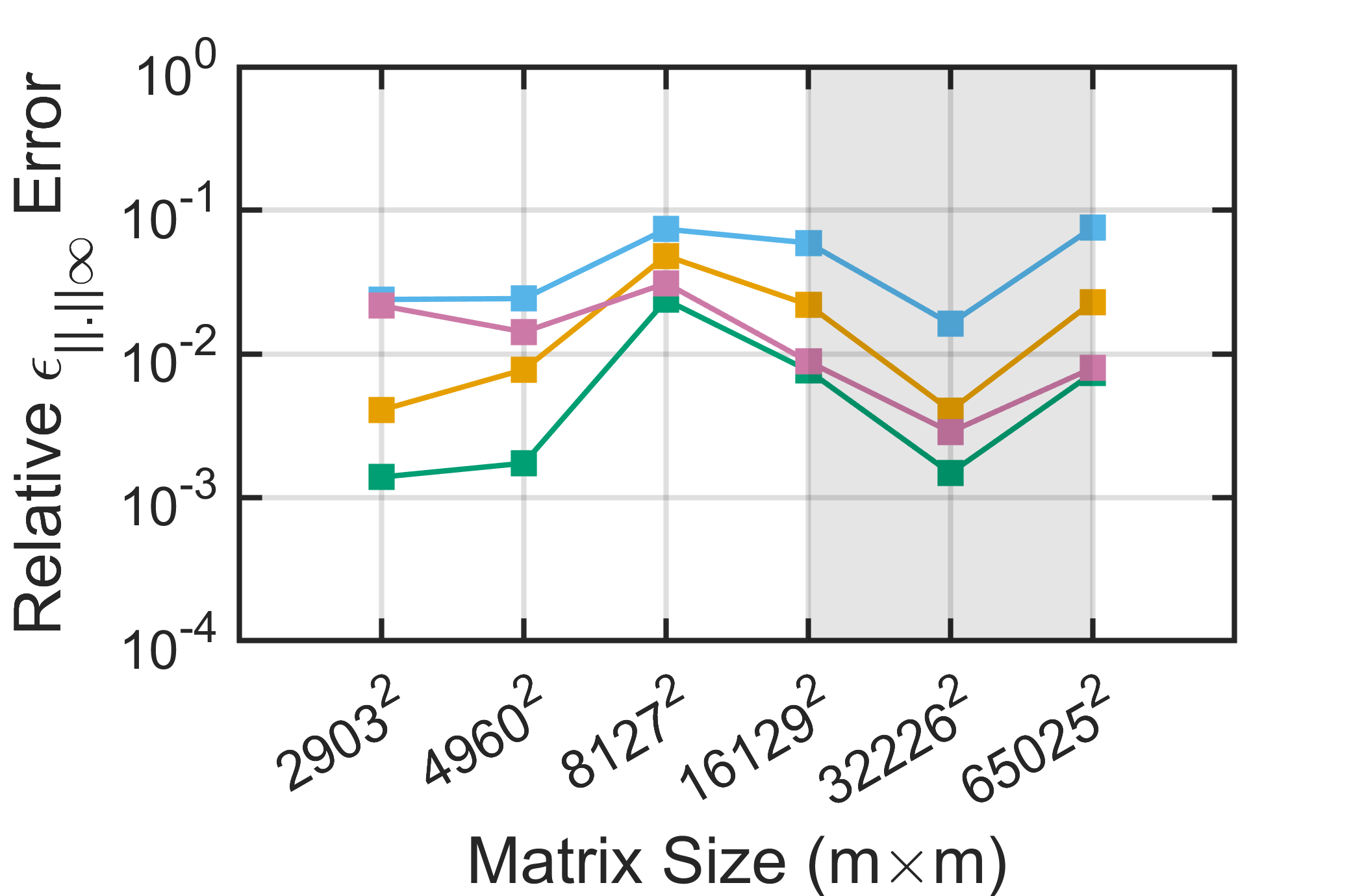}}
        \end{minipage}
    
        \begin{minipage}[t]{0.49\textwidth}
            \centering
            \subfigure[]{\includegraphics[width=\textwidth]{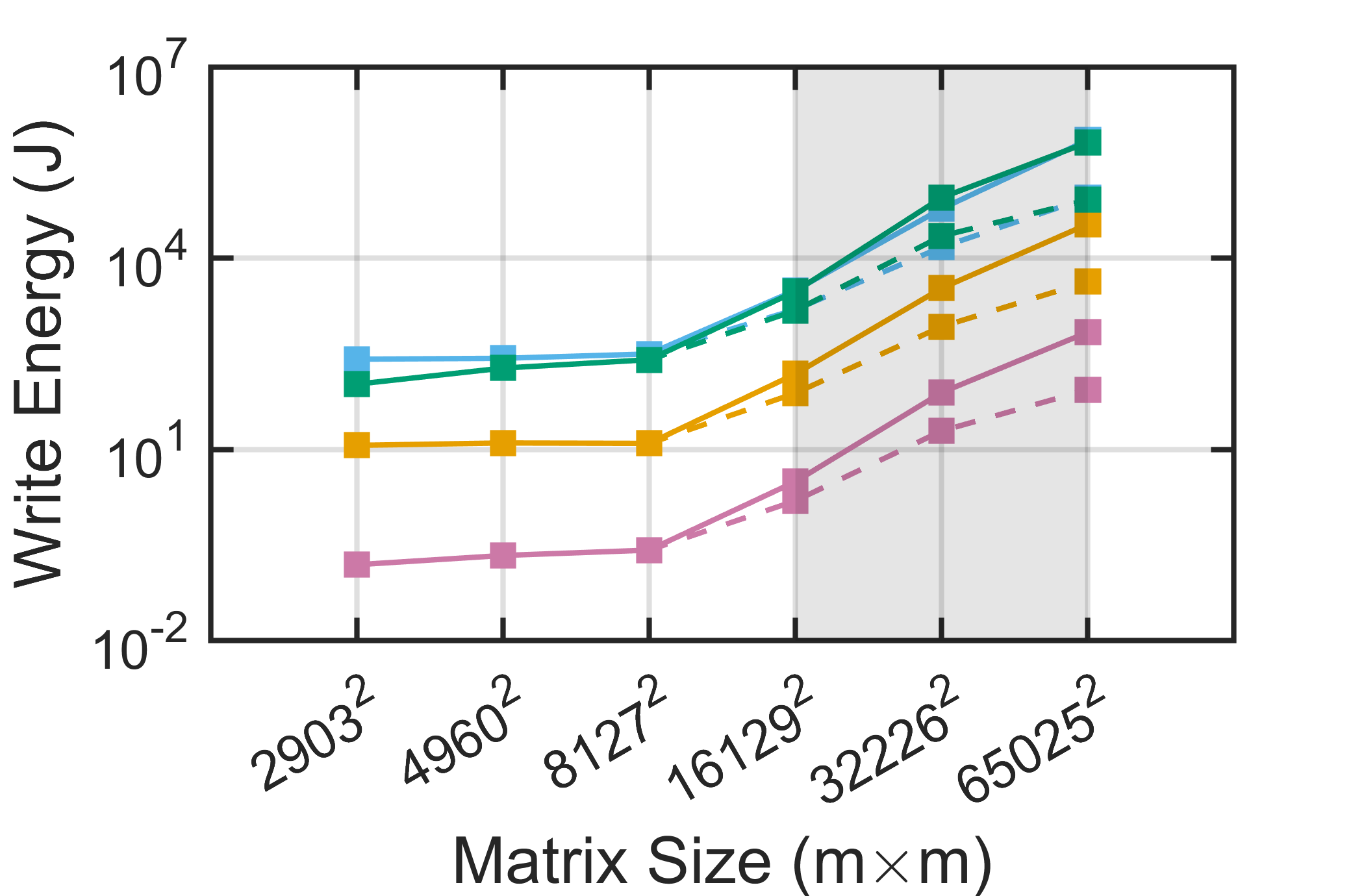}}
        \end{minipage}
        \hfill
        \begin{minipage}[t]{0.49\textwidth}
            \centering
            \subfigure[]{\includegraphics[width=\textwidth]{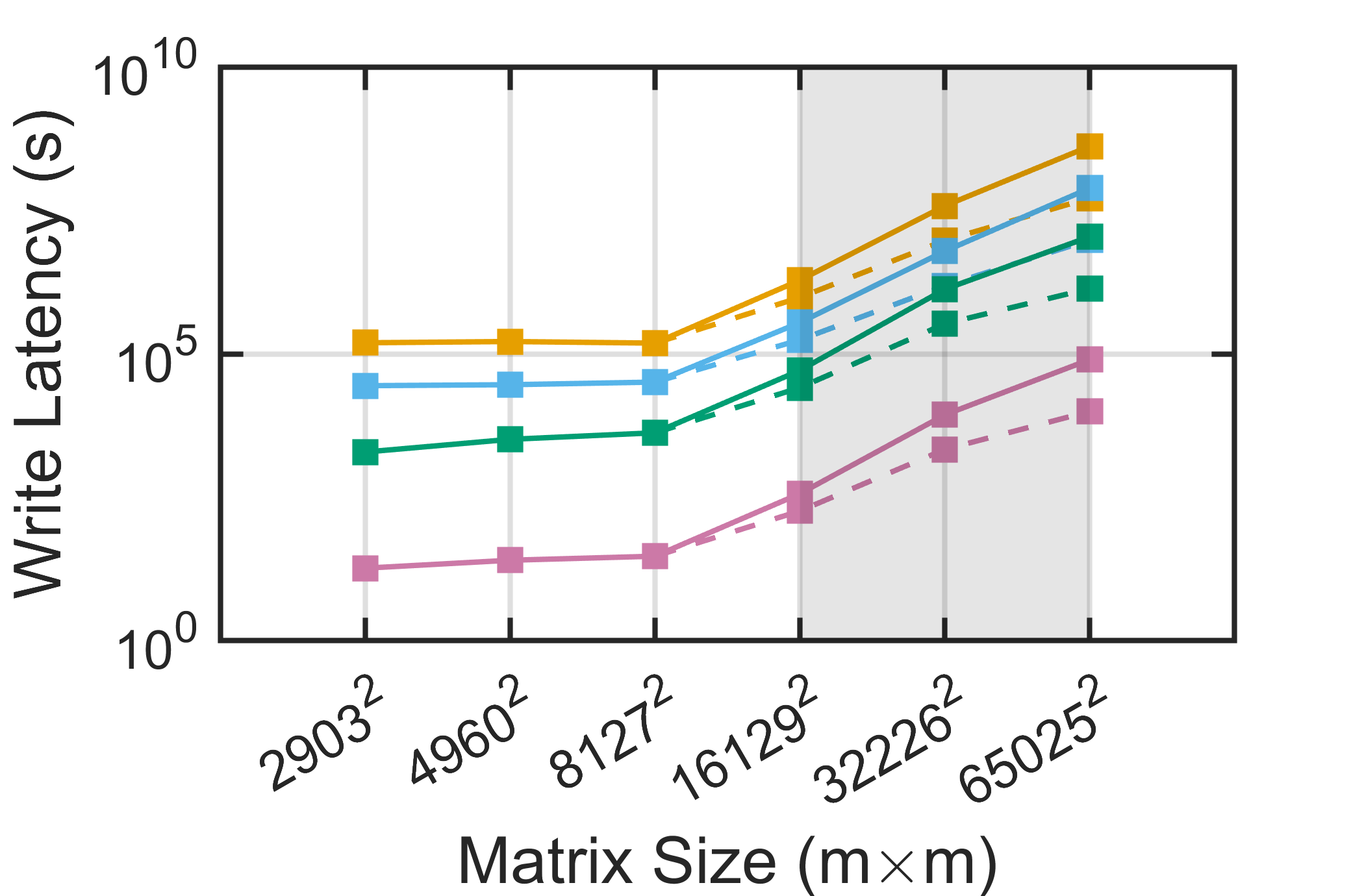}}
        \end{minipage}
    
    \caption{A comparison of device performance across key metrics: (a) Relative $\ell_2$-norm Error, (b) Relative $\ell_\infty$-norm Error, (c) Write Energy, (d) Write Latency. The performance is evaluated on different problem sizes using a multi-MCA system with a fixed tile size and cell sizes per MCA. Due to the concurrent nature of distributing the MVM workload across a multi-MCA system of $8\times8$ tile sizes, write energy and write latency values were computed as the mean across all MCA devices. Furthermore, due to virtualization, a normalizing factor (dashed lines) is applied starting from the $16129\times16129$ matrix size.}\label{fig:strongScaling_results} 
\end{figure*}

Fig.~\ref{fig:strongScaling_results} presents the performance results of the systems simulated by MELISO+ in these strong scaling experiments, where each point on the x-axis representing a distinct matrix obtained from the aforementioned matrix collection that broadly aligns with an intended doubling ($2\times$) in scale.
To ensure a fair comparison across varying matrix sizes, we introduced a normalization factor based on the number of reassignments that a single MCA would undergo due to virtualization, particularly for matrices ranging from $16{,}129\times16{,}129$ to $65{,}025\times65{,}025$. As discussed in the previous section, virtualization arises when the system's physical dimensions do not match the problem size, causing each MCA to be assigned multiple times. For instance, consider the \textbf{Dubcova1} matrix used in these experiments—whose size is $16{,}129\times16{,}129$—where the system's physical dimensions are $(8\times1024) \times (8\times1024)$. In this case, each MCA is assigned approximately two (2) times; thus, the normalization constant is two (2). This adjustment ensures that both $\text{E}_{\text{w}}$ and $\text{L}_{\text{w}}$ are accurately represented. 

Results in Fig.~\ref{fig:strongScaling_results} mirror those described in the previous section: when our proposed methods are implemented, the simulated systems consistently showcase low relative error norms, underscoring the robustness and efficacy of MELISO+ over a wide spectrum of problem sizes.
The trends in Fig.~\ref{fig:strongScaling_results} reveal some volatility in relative errors across all device types, which is attributed to inherent properties of the selected matrices, such as condition number, matrix norms, and sparsity patterns. Despite these variations, Fig.~\ref{fig:strongScaling_results} confirms that the relative performance of devices remains consistent and aligns with expected behavior, which showcases the reliability of MELISO+ in handling diverse matrix characteristics.
Notably, virtualization has a more prominent impact in the strong-scaling regime compared to the weak-scaling regime (as previously seen in Fig.~\ref{fig:commercialized_results}), as both $\text{E}_{\text{w}}$ and $\text{L}_{\text{w}}$ increase exponentially as the problem size increases. 

\section{Discussion}\label{sec13}


The proposed MELISO+ framework is a comprehensive end-to-end benchmarking tool designed to evaluate error propagation in linear computing operations, tailored for in-memory systems based on RRAM devices. The framework enables systematic analysis of how device-level metrics and algorithmic design considerations impact error magnitude and distribution, while also supporting distributed computing at scale and the integration of novel, hardware-aware error correction techniques. 

Our framework addresses two fundamental barriers that currently limit the commercial-scale adoption of RRAM technology by integrating the following key innovations:
i) Distributed In-Memory Computing for Scalability:
We introduce a distributed in-memory computing architecture that leverages virtualization to orchestrate large-scale linear operations across multiple RRAM-based MCAs. This architecture is inherently scalable and enables the execution of realistic, high-dimensional workloads that exceed the capacity of individual MCAs, thereby meeting the computational demands of modern applications such as machine learning and generative AI. ii) Error Correction Algorithm to Mitigate Device-Level Non-Idealities:
We develop a first-of-its-kind error correction algorithm, purpose-built for RRAM-based in-memory computing. Unlike digital correction schemes, this algorithm is designed from the ground up to mitigate first-order errors arising from RRAM device non-idealities, such as limited memory windows and cycle-to-cycle variability. MELISO+ enables reliable and energy-efficient computation even when operating on RRAM devices with suboptimal characteristics, thus extending the practical usability of lower-cost or lower-precision hardware.

A key fundamental insight highlighted by our study is that scalable linear operations can be achieved for in-memory devices by leveraging a distributed and massively parallel architecture comprising many MCAs. Our results show that our proposed framework can scale to accommodate in-memory compute even for matrices with dimensions up to $65{,}000\times65{,}000$ in a computationally reliable fashion.

Our proposed first-of-its-kind two-level error-correction algorithm ensures that a low-accuracy, low-latency MCA (for instance, TaO\textsubscript{x}-HfO\textsubscript{x} with a limited memory window) can outperform a high-accuracy, high-latency MCA (such as EpiRAM with a higher memory window), resulting in the same level of accuracy and improvements of 3-4 orders of magnitude in energy use and 2 orders of magnitude in latency.


These advances establish MELISO+ as a robust foundation for evaluating and scaling in-memory computing hardware, while also underscoring the transformative impact of algorithm–hardware co-design in overcoming fundamental device limitations.

\section{Methods}\label{sec:Methods}

In this section, we describe our experiment setup and the key components of MELISO+. These components include (i) first-order error correction, (ii) second-order error correction, and (iii) the distributed paradigm for large-scale computations.

\subsection{Experimental Setup}\label{subsec4.1}
We conducted all our experiments on Oklahoma State University's (OSU) Pete High Performance Computing infrastructure. We utilized the OpenMPI library~\cite{gabriel2004open} along with its Python bindings, provided by~\texttt{mpi4py}~\cite{dalcin2021mpi4py}, to enable large-scale distributed processing.

\subsection{First-order Error Correction}\label{subsec3.1}
Let $\mathbf{A} \in \mathbb{R}^{m\times n}$ be the input matrix and $\mathbf{x} \in \mathbb{R}^{n\times 1
}$ be the input vector. Let $\mathbf{a}_j$ be an arbitrary column vector of $\mathbf{A}$ where $1\leq j \leq n$, and let $\mathbf{a}^\top_i$ be an arbitrary row vector of $\mathbf{A}$ where $1\leq i \leq m$. Thus, the input matrix $\mathbf{A}$ can be described with the column and row vectors as:
\begin{equation}
    \mathbf{A} = \begin{bmatrix}
        \rotvert \mathbf{a}_1^\top \rotvert \\
        \rotvert \mathbf{a}_2^\top \rotvert \\
        \vdots \\
        \rotvert \mathbf{a}_i^\top \rotvert \\
        \vdots \\
        \rotvert \mathbf{a}_m^\top \rotvert \\
        \end{bmatrix}
         = \begin{bmatrix}
         \vert & \vert & & \vert & \vert \\
         \mathbf{a}_1 & \mathbf{a}_2 & \cdots & \mathbf{a}_j &  \mathbf{a}_n \\
         \vert & \vert & & \vert & \vert \\
         \end{bmatrix}
\end{equation}

When assigned to an MCA, the matrix $\mathbf{A}$ and the vector $\mathbf{x}$ are subject to mapping errors due to the stochastic nature of the device. Let $\mathbf{\tilde{A}}$ and $\mathbf{\tilde{x}}\in\mathbb{R}^{r}$ represent the encoded matrix and vector on an MCA device, respectively. Thus, the relationship between the true objects (i.e., $\mathbf{A}$ and $\mathbf{x}$) and their respective encoded counterparts (i.e.,  $\mathbf{\tilde{A}}$ and $\mathbf{\tilde{x}}$) is derived as follows: 

\begin{equation}
    \mathbf{\tilde{x}} = \mathcal{H}(\mathbf{x}) = \mathbf{x}\big(1 + \pmb{\epsilon}_{\mathbf{x}}\big) 
\end{equation}

\begin{equation}
    \begin{aligned}
        \mathbf{\tilde{A}} &= \mathcal{H}(\mathbf{A}) = \mathbf{A}\big(1 + \pmb{\epsilon}_{\mathbf{A}}\big)\\
        &= 
        \begin{bmatrix}
        \rotvert \mathbf{a}_1^\top\big(1 + \pmb{\epsilon}_{\mathbf{a}_1^\top}\big) \rotvert \\
        \rotvert \mathbf{a}_2^\top\big(1 + \pmb{\epsilon}_{\mathbf{a}_2^\top}\big) \rotvert \\
        \vdots \\
        \rotvert \mathbf{a}_2^\top\big(1 + \pmb{\epsilon}_{\mathbf{a}_2^\top}\big) \rotvert 
        \end{bmatrix}
    \end{aligned}
\end{equation}

where $\mathcal{H}(\bullet)$ is a function that programs the real values of $\mathbf{A}$ and $\mathbf{x}$ to their corresponding synaptic weights in the MCA, and $\pmb{\epsilon}_{\bullet}$ is the programming error induced by the stochasticity of the MCA device. In MELISO+, the assignments of the input matrix and input vector are performed by the \texttt{adjustableMatWriteandVerify} and \texttt{adjustableVecWriteandVerify} functions, respectively. In these functions, the circuit-level assignment of matrix or vector elements is conducted by the $\texttt{MCAsetWeights}$ protocol. 

\begin{algorithm}[h]
    \caption{adjustableMatWriteandVerify}\label{alg:adjustableMatWriteandVerify}
    \begin{algorithmic}[1]
    \Require $\mathbf{A}\in\mathbb{R}^{m\times n}$; $\epsilon\in\mathbb{R}$; $N\in\mathbb{Z}^+$; $p\in\{2,\infty\}$.
    \Ensure $\mathbf{\tilde{A}}^{\{k\}} \in \mathbb{R}^{m\times n}$, $k\in\mathbb{Z}^+$, $\delta(\mathbf{A},\mathbf{\tilde{A}}^{\{k\}}) \in \mathbb{R}$. 
    \State $k \leftarrow 0$;
    \State $\mathbf{\tilde{A}}^{\{k\}}$ $\leftarrow$~\texttt{MCAsetWeights}($\mathbf{A}$);
    \While{$(k < N)$ \textbf{and} $\big[\delta(\mathbf{A},\mathbf{\tilde{A}}^{\{k\}}) > \epsilon\big]$}:
        \State $k$ $\leftarrow$ $k + 1$;
        \State $\delta(\mathbf{\tilde{A}}^{\{j\}}, \mathbf{A})$ $\leftarrow$ $||\mathbf{\tilde{A}}^{\{j\}} - \mathbf{A}||_p$;
        \If {$\delta(\mathbf{\tilde{A}}^{\{j\}}, \mathbf{A}) > \pmb{\epsilon}$}: \\
         \quad $\mathbf{\tilde{A}}^{\{j\}}$ $\leftarrow$~\texttt{MCAsetWeights}($\mathbf{A}$);
        \EndIf
    \EndWhile
    \end{algorithmic}\label{alg:adjustableMatWriteandVerify}
\end{algorithm}

\begin{algorithm}[h]
\caption{adjustableVecWriteandVerify}\label{alg:adjustableVecWriteandVerify}
\begin{algorithmic}[1]
\Require $\mathbf{x}\in\mathbb{R}^{n\times1}$; $\epsilon\in\mathbb{R}$; $N\in\mathbb{Z}^+$; $p\in\{2,\infty\}$.
\Ensure $\mathbf{\tilde{x}}^{\{k\}}\in\mathbb{R}^{n\times1}$; $k\in\mathbb{Z}^+$; $\delta(\mathbf{x},\mathbf{\tilde{x}}^{\{k\}})\in\mathbb{R}$.
\State $k \leftarrow 0$;
\State $\mathbf{\tilde{x}}^{\{k\}}$ $\leftarrow$~\texttt{MCAsetWeights}($\mathbf{x}$);
\While{$(k < N)$ \textbf{and} $\big[\delta(\mathbf{x},\mathbf{\tilde{x}}^{\{k\}}) > \epsilon\big]$}:
    \State $k$ $\leftarrow$ $k + 1$;
    \State $\delta(\mathbf{\tilde{x}}^{\{k\}}, \mathbf{x})$ $\leftarrow$ $||\mathbf{\tilde{x}}^{\{k\}} - \mathbf{x}||_p$;
    \If {$\delta(\mathbf{\tilde{x}}^{\{k\}}, \mathbf{x}) > \epsilon$}: \\
     \quad $\mathbf{\tilde{x}}^{\{k\}}$ $\leftarrow$~\texttt{MCAsetWeights}($\mathbf{x}$);
    \EndIf
\EndWhile
\end{algorithmic}\label{alg:adjustableVecWriteandVerify}
\end{algorithm}

To correct the induced errors, we derive three matrix-vector products: 

\begin{itemize}
    \item The product of the original matrix $\mathbf{A}$ and the synaptic vector $\mathbf{\tilde{x}}$, denoted by $\mathbf{\tilde{u}} \in \mathbb{R}^{m\times1}$:
        \begin{equation}
        \begin{aligned}
            u_i &= \mathbf{a}_i^\top\mathbf{x}\big(1+\pmb{\epsilon}_{\mathbf{x}}\big), &1\leq i \leq m\\
        \end{aligned}
    \end{equation}

    \item The product of the synaptic matrix $\mathbf{\tilde{A}}$ and the original vector $\mathbf{x}$, denoted by $\mathbf{\tilde{v}} \in \mathbb{R}^{m\times1}$:
        \begin{equation}
            \begin{aligned}
                \tilde{v}_i &= \mathbf{\tilde{a}}_i^\top \mathbf{x}, &1\leq i \leq m\\
                &= \mathbf{a}_i^\top\mathbf{x}\big(1+\pmb{\epsilon}_{\mathbf{a}_i^\top}\big), &1\leq i \leq m\\
            \end{aligned}
        \end{equation}
    \item Finally, the product of the synaptic matrix $\mathbf{\tilde{A}}$ and the synaptic vector $\mathbf{\tilde{x}}$, denoted by $\mathbf{\tilde{y}} \in \mathbb{R}^m$: 
        \begin{equation}
        \begin{aligned}
            \tilde{y}_i &= \mathbf{\tilde{a}}_i^\top \mathbf{\tilde{x}}, &1\leq i \leq m \\
            &= \mathbf{a}_i^\top\mathbf{x}\big(1+\pmb{\epsilon}_{\mathbf{x}}\big)\big(1+\pmb{\epsilon}_{\mathbf{a}_i^\top}\big), &1\leq i \leq m \\
            &= \mathbf{a}_i^\top\mathbf{x}(1 + \pmb{\epsilon}_{\mathbf{a}_i^\top} + \pmb{\epsilon}_{\mathbf{x}} + \pmb{\epsilon}_{\mathbf{a}_i^\top}\pmb{\epsilon}_{\mathbf{x}}), &1\leq i \leq m \\
        \end{aligned}
    \end{equation}
\end{itemize}

Let $\mathbf{p}\in\mathbb{R}^{m\times1}$ be a linear combination of these products s.t.: $\mathbf{p} = \mathbf{\tilde{v}} - \mathbf{\tilde{y}} + \mathbf{\tilde{u}}$. Thus, each element of $\mathbf{p}$ is given as:

\begin{equation}
    \begin{aligned}
    &p_i \\
    &= \mathbf{a}_i^\top\mathbf{x}\big(1+\pmb{\epsilon}_{\mathbf{a}_i^\top}\big) - \mathbf{a}_i^\top\mathbf{x}\big(1+\pmb{\epsilon}_{\mathbf{x}}\big)\big(1+\pmb{\epsilon}_{\mathbf{a}_i^\top}\big) + \mathbf{a}_i^\top\mathbf{x}\big(1+\pmb{\epsilon}_{\mathbf{x}}\big) \\
    &= \mathbf{a}_i^\top\mathbf{x}\Big(1+\pmb{\epsilon}_{\mathbf{a}_i^\top} + 1+\pmb{\epsilon}_{\mathbf{x}} -
    1 - \pmb{\epsilon}_{\mathbf{a}_i^\top} - \pmb{\epsilon}_{\mathbf{x}} - \pmb{\epsilon}_{\mathbf{a}_i^\top}\pmb{\epsilon}_{\mathbf{x}} \Big)\\
    &= \mathbf{a}_i^\top\mathbf{x}\big(1 - \pmb{\epsilon}_{\mathbf{a}_i^\top}\pmb{\epsilon}_{\mathbf{x}}\big), \quad 1 \leq i \leq m
    \end{aligned}\label{equation:FOerrorCorrection}
\end{equation}

Indisputably, this linear combination eliminates the first-order error terms, leaving only second-order errors that will be addressed in the following component.

\subsection{Second-order Error Correction}\label{subsec3.2}

As concluded in the description of the previous component, the vector $\mathbf{p}$ still contains second-order errors. Let $\mathbf{b}\in\mathbb{R}^{m\times1}$ be the true MVM result—that is, our ground truth, where each element of $\mathbf{b}$ is $b_i = \mathbf{a}_i^\top\mathbf{x}$ for all $1\leq i \leq m$ and is unknown to us. Thus, we can describe the relationship between $\mathbf{p}$ and $\mathbf{b}$ as:

\begin{equation}
    \mathbf{p} = \mathbf{b} + \pmb{\epsilon}
\end{equation}

In other words, $\mathbf{p}$ is the noisy measurement of $\mathbf{b}$. with $\pmb{\epsilon} \in \mathbb{R}^{m\times1}$ is the noise vector representing the second-order errors. To minimize the impact of this noise, we formulate a regularized least-squares problem:

\begin{equation*}
    \min_{\mathbf{p}} \quad \|\mathbf{b} - \mathbf{p}\|_2^2 + \lambda \|\mathbf{L} \mathbf{\mathbf{p}}\|_2^2
\end{equation*}

where $\mathbf{L} \in \mathbb{R}^{n\times n}$ is the first-order differential square matrix, defined as:

\begin{equation}
    \mathbf{L} := 
    \begin{cases}
    1; & \text{if } i = j \\
    h\in\mathbb{Z}^-; & \text{if } j = i + 1, \\
    0; & \text{otherwise}
    \end{cases}
\end{equation}

for all $1\leq i,j \leq n$ and $h=-1$ in many denoising applications~\cite{selesnick2013least}. The closed-form solution to this problem is given as:

\begin{equation}
    \mathbf{y}(\lambda) = (\mathbf{I}_n + \lambda \mathbf{L}^\top \mathbf{L})^{-1} \mathbf{p}
\end{equation}

where $\lambda \in (0,\infty)$ is a regularization parameter, and $\mathbf{I}_n$ is an identity matrix of size $n\times n$. Since the noise vector $\pmb{\epsilon}$ is negative, as seen in Eq.~\ref{equation:FOerrorCorrection}, by pre-conditioning $\lambda$ within the $(0,1)$ interval, the second-order term is exponentially attenuated. In MELISO+, the regularized least squares denoising is conducted by the~\texttt{denoiseLeastSquare} function, while the MVM with both first- and second-order error correction schemes is implemented by the~\texttt{correctedMatVecMul} function. These functions are described in Supplementary Information~\ref{secB1.1}. and~\ref{secB1.2}

\subsection{Distributed Paradigm}\label{subsec3.3}
To execute MVM operations with large-scale matrices that exceed the physical limits of a single MCA device, we propose a distributed paradigm using a multi-chiplet configuration with each single chiplet representing an MCA device. This configuration consists of an $R \times C$ array of MCA devices (where $R \geq C$), and each MCA contains a $r \times c$ number of cells, where $r \geq c$. With this configuration, we can accommodate matrices of size up to $(R \times r) \times (C \times c)$. For larger matrices, we introduce a novel virtualization technique comprising two primary steps: Dimension Matching, and Chunk Partitioning. 



In the first step, there are three types of scenarios that can happen:
\begin{itemize}
    \item \textbf{Ideal Case}:  Let $p$ and $q$ represent the respective row and column indices for each MCA device s.t. $1 \leq p \leq R$ and $1 \leq q \leq C$. Similarly, let $l$ and $h$ represent the respective row and column indices for each cell in a single MCA device, such that $1 \leq l \leq R$ and $1 \leq h \leq c$. We define an ideal case as when the size of the input matrix $\mathbf{A}$ matches the
    system dimensions (i.e.,  $m = R \times r$ and $n = C \times c$). In this case, each element $a_{ij}$ of $\mathbf{A}$ is mapped to each cell of the system, denoted by $\tilde{a}_{pqlh}$.
    \item \textbf{Non-ideal Case}: In a non-ideal case, the size of at least one dimension of the input matrix $\mathbf{A}$ is smaller than the system's corresponding dimension (i.e., $m < R\times r$ or $n <C\times c$). Thus, we use the zero-padding technique to match the dimensions, which is executed by the function~\texttt{zeroPadding}, described in Supplementary Information~\ref{secB1.3}. Specifically, we either add $(R\times r - m)$ zero rows or $(C\times c - n)$ zero columns. 
    
    \item \textbf{Large-Scale Matrices}: For matrices of sizes that are significantly larger than the system dimensions (e.g., $m \gg R \times r$ and $n \gg C \times r$), we divide the matrix into smaller blocks, denoted by $\mathbf{A}^{(i)}$ using the~\texttt{blockPartition} function. Each block is then processed independently, and zero-padding is applied to the block if necessary to match the system's physical dimensions.
    
        \begin{algorithm}[h]
            \caption{blockPartition}\label{alg:blockPartition}
            \begin{algorithmic}[1]
                \Require $\mathbf{A}\in\mathbb{R}^{m\times n}; R, r, C,c \in\mathbb{Z}^{+}$. 
                \Ensure A collection of matrix blocks: $\mathcal{B}_{\mathbf{A}} = \{\mathbf{A}^{(i)}\}$.
                \State $\mathcal{B}_{\mathbf{A}}\leftarrow \{\emptyset\}$;
                \State $m, n \leftarrow \text{shape}(\mathbf{A})$;
                \State $\hat{m} \leftarrow \lceil m /(R\times r) \rceil$; 
                \State $\hat{n} \leftarrow \lceil n /(C\times c) \rceil$;
                \For{$1 \leq i\leq \hat{m}$}
                    \For{$1 \leq j \leq \hat{n}$}
                        \State $\mathbf{\hat{A}}_{[i,j]} \leftarrow~\texttt{indexing}(\mathbf{A})$
                        \State $ \mathcal{B}_{\mathbf{A}} \leftarrow \mathcal{B}_{\mathbf{A}} \cup \{\mathbf{\hat{A}}_{(i,j)}\}$ 
                    \EndFor
                \EndFor
            \end{algorithmic}
        \end{algorithm}
\end{itemize}

Moving to the second step, once the dimensions are matched, the input matrix $\mathbf{A}$ (in the first two scenarios) or a single block $\mathbf{A}^{(i)}$ (in the last scenario) is partitioned into a $R\times C$ number of smaller chunks using the function~\texttt{generateMatChunksSet} (see Supplementary Information~\ref{secB1.3}). Similarly, the input vector $\mathbf{x}$ is also partitioned into an appropriate number of chunks to match those from the matrices. These matrix-vector pairs of chunks are then assigned to individual MCA devices for parallel computation, conducted by the~\texttt{distributedMatVecMul} function.

\begin{algorithm}[h]
\caption{distributedMatVecMul}\label{alg:distributedMatVecMul}
\begin{algorithmic}[1]
\Require $\mathbf{A}\in\mathbb{R}^{m\times n}$; $\mathbf{x}\in\mathbb{R}^{n}$; $R,C,r,c\in\mathbb{Z}^+$
\Ensure $\mathbf{b}\in\mathbb{R}^{n\times1}$
\State $m, n \leftarrow \text{shape}(\mathbf{A})$;
\State $\texttt{zeroPadding}(R,C,r,c)$;
\State $\mathcal{S}_{\mathbf{A}} \leftarrow~\texttt{generateMatChunksSet}(\mathbf{A})$
\State $\mathcal{S}_{\mathbf{x}} \leftarrow~\texttt{generateVecChunksSet}(\mathbf{x})$
\State $\mathcal{S}_{\mathbf{b}} \leftarrow \{\emptyset\}$\\
\text{\textbf{do in parallel} for each chunk }:
    \State \quad $\mathbf{A}_{\text{chunk}} \leftarrow \text{indexing}(\mathcal{S}_{\mathbf{A}})$;
    \State \quad $\mathbf{x}_{\text{chunk}} \leftarrow \text{indexing}(\mathcal{S}_{\mathbf{b}})$;
    \State \quad $\mathbf{b}_{\text{chunk}} \leftarrow~\texttt{correctedMatVecMul}(\mathbf{A},\mathbf{x})$
    \State \quad $\mathbf{b}_{\text{chunk}} \leftarrow S_b \cup \mathbf{b}_{\text{chunk}} $
\State $\mathbf{b} \leftarrow~\texttt{vectorize}(\mathcal{S}_{\mathbf{b}})$
\end{algorithmic}
\end{algorithm}

\backmatter

\section*{Acknowledgements}

\section*{Declarations}
The authors declare no competing interests.

\bibliography{sn-bibliography}









\newpage

\section*{Supplementary information}

\appendix

\section{Employed matrices}\label{secA1}
Table~\ref{tabular:MatrixSummary} provides the summary of squared matrices employed in our experiments pertaining to MELISO+, obtained from the SuiteSparse Matrix Collection \cite{Kolodziej2019}. In Table~\ref{tabular:MatrixSummary}, 

\begin{itemize}
    \item \textbf{nzeroes} represents the percentage of matrix entries that have a value of zero.
    \item \textbf{Pattern Symmetry} represents the percentage of nonzero entries that have a matching nonzero entry across the diagonal.
    \item \textbf{Numeric Symmetry} represents the percentage of nonzero entries that are numerically symmetric.
    \item $||\mathbf{A}||_2$ is the matrix norm (i.e., the spectral norm), and $\kappa(\mathbf{A})$ is the matrix conditioning number.
\end{itemize}

\begin{table}[h]
\centering
\begin{tabular}{lccccccc}
\multicolumn{1}{c}{\textbf{Matrix}} & \multicolumn{1}{c}{dim($\mathbf{A}$)} & \multicolumn{1}{c}{\textbf{\begin{tabular}[c]{@{}c@{}}nzeros\\ (\%)\end{tabular}}} & \multicolumn{1}{c}{\textbf{\begin{tabular}[c]{@{}c@{}}Pattern\\ Symmetry\\ (\%)\end{tabular}}} & \multicolumn{1}{c}{\textbf{\begin{tabular}[c]{@{}c@{}}Numeric\\ Symmetry\\ (\%)\end{tabular}}} & \multicolumn{1}{c}{$||\mathbf{A}||_2$} & \multicolumn{1}{c}{$\kappa$(\textbf{A})} & \multicolumn{1}{c}{\textbf{Results}} \\ \hline
bcsstk02                            & 66                               & 0                                                                               & 100                                                                                            & 100                                                                                            & 1.822575e+04                       & 4.324971e+03                       & \ref{subsection:4.2}                                   \\
wang2                               & 2,903                            & 0                                                                               & 100                                                                                            & 80.8                                                                                           & 4.138078                           & 2.305543e+04                       & \ref{subsubsection:Strong-Scaling}                                   \\
add32                               & 4,960                            & 1.6898                                                                          & 100                                                                                            & 30.5                                                                                           & 5.749318e-02                       & 1.366769e+02                       & \ref{subsubsection:Weak-Scaling}, \ref{subsubsection:Strong-Scaling}                                  \\
c-38                                & 8,127                            & 0                                                                               & 100                                                                                            & 100                                                                                            & 6.083484e+02                       & 1.530683e+04                       & \ref{subsubsection:Strong-Scaling}                                   \\
Dubcova1                            & 16,129                           & 0                                                                               & 100                                                                                            & 100                                                                                            & 4.796329                           & 9.971199                           & \ref{subsubsection:Strong-Scaling}                                   \\
helm3d01                            & 32,226                           & 0                                                                               & 100                                                                                            & 100                                                                                            & 5.052177e-01                       & 2.451897e+05                       & \ref{subsubsection:Strong-Scaling}                                   \\
Dubcova2                            & 65,025                           & 0                                                                               & 100                                                                                            & 100                                                                                            & *                        & *                         & \ref{subsubsection:Strong-Scaling}                                   \\ \hline
\multicolumn{8}{l}{\begin{tabular}[c]{@{}l@{}}*: These information are not provided in the SuiteSparse Matrix Collection.\end{tabular}} 
\end{tabular}
\caption{Summary of distinct matrices.}\label{tabular:MatrixSummary}
\end{table}

\section{Benchmarking the error correction method with \textbf{bcsstk02}}\label{secA2}
Figs.~\ref{fig:bcsstk02_EC_0} and~\ref{fig:bcsstk02_EC_1} illustrate trends in these performance metrics, acquired from the MVM with \textbf{bcsstk02}, as the iteration count in the~\texttt{adjustableWriteandVerify} function increases up to $k=20$. Similar to Fig.~\ref{fig:Iperturb_EC_0} and Fig.~\ref{fig:Iperturb_EC_1}, each data point in these figures represents the average of 100 experimental replicates, ensuring the statistical reliability of the observed trends. 

In Fig.~\ref{fig:bcsstk02_EC_0}, relative error norms decrease exponentially with increasing iteration count for most devices. However, for EpiRAM devices, which are inherently high-performance, applying multiple iterations ($k > 0$) introduces additional inaccuracies, negatively impacting the final results.
Conversely, Fig.~\ref{fig:bcsstk02_EC_1} shows that combining the proposed error correction methods with the multi-iteration~\texttt{adjustableWriteandVerify} scheme results in a more significant decrease in relative error norms compared to Fig.~\ref{fig:bcsstk02_EC_0}. This decrease stabilizes after approximately five iterations for most materials, suggesting that $k=5$ is sufficient for optimal performance.
Notably, our proposed methods significantly enhance the performance of AlO\textsubscript{x}-HfO\textsubscript{2} and TaO\textsubscript{x}-HfO\textsubscript{x} devices, previously considered among the lower performing RRAM devices for accuracy~\cite{10.1145/3354265.3354266}. These improvements enable them to achieve statistically equivalent performance to EpiRAM, as detailed in Table~\ref{tabular:errorCorrection&setWeightsIncremental}. 
\setcounter{figure}{0}

\begin{figure*}
   \makeatletter
    \renewcommand{\thefigure}{S\@arabic\c@figure}
    \makeatother
    \centering
    \begin{minipage}[t]{0.49\textwidth}
        \centering
        \subfigure[]{%
            \includegraphics[width=\textwidth]{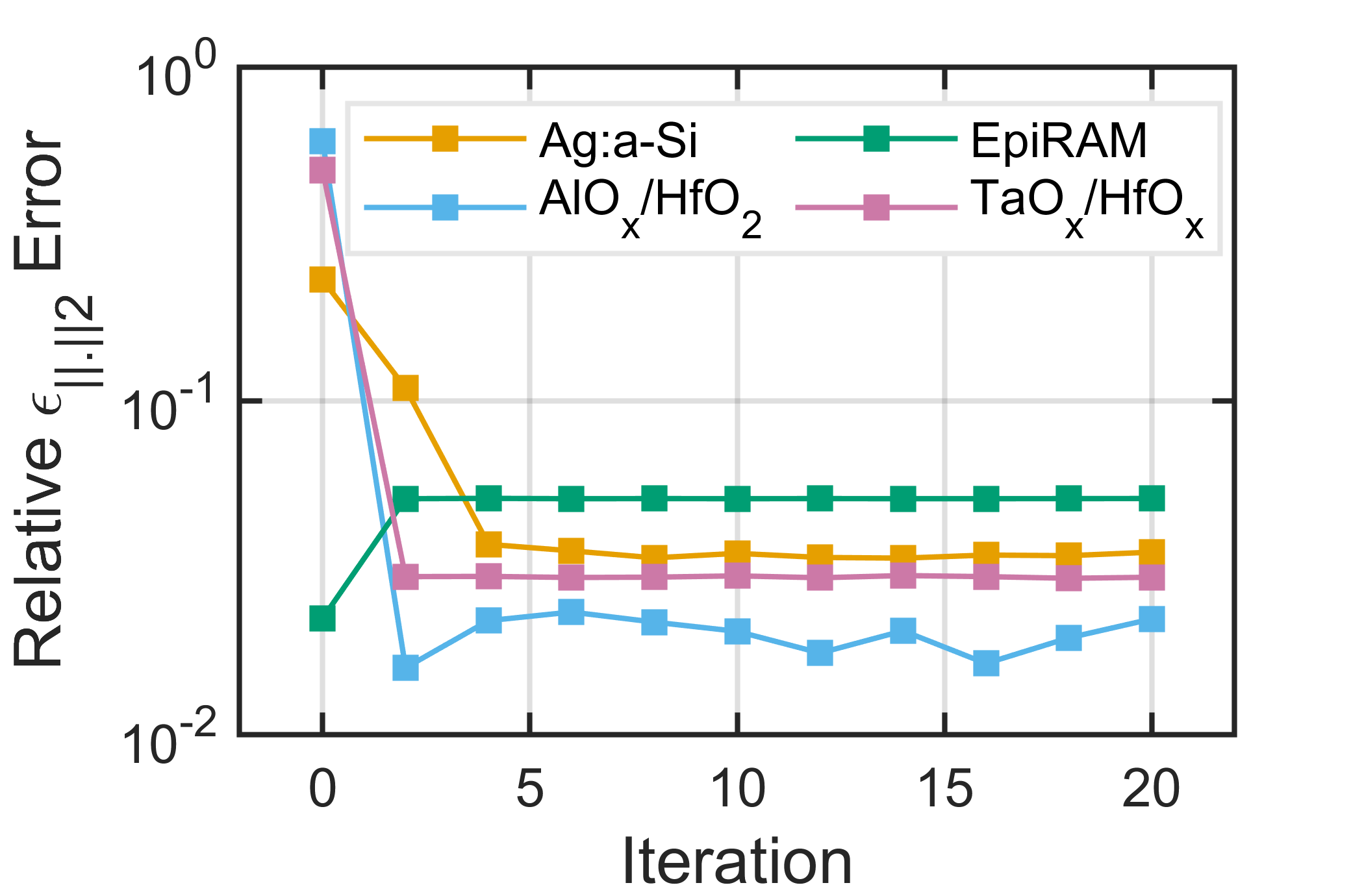}%
        }
    \end{minipage}
    \hfill
    \begin{minipage}[t]{0.49\textwidth}
        \centering
        \subfigure[]{%
            \includegraphics[width=\textwidth]{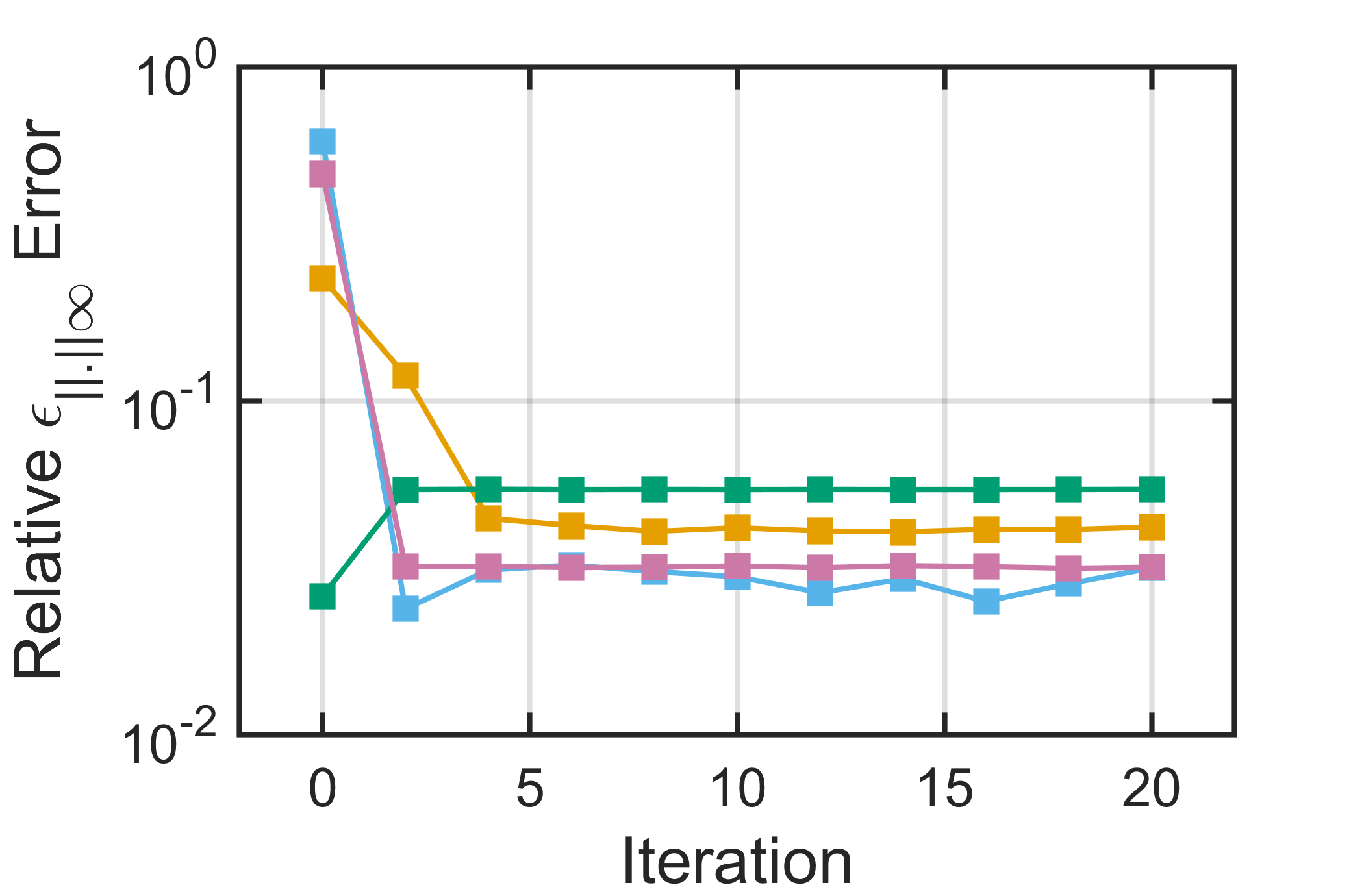}%
        }
    \end{minipage}
    
    \begin{minipage}[t]{0.49\textwidth}
        \centering
        \subfigure[]{%
            \includegraphics[width=\textwidth]{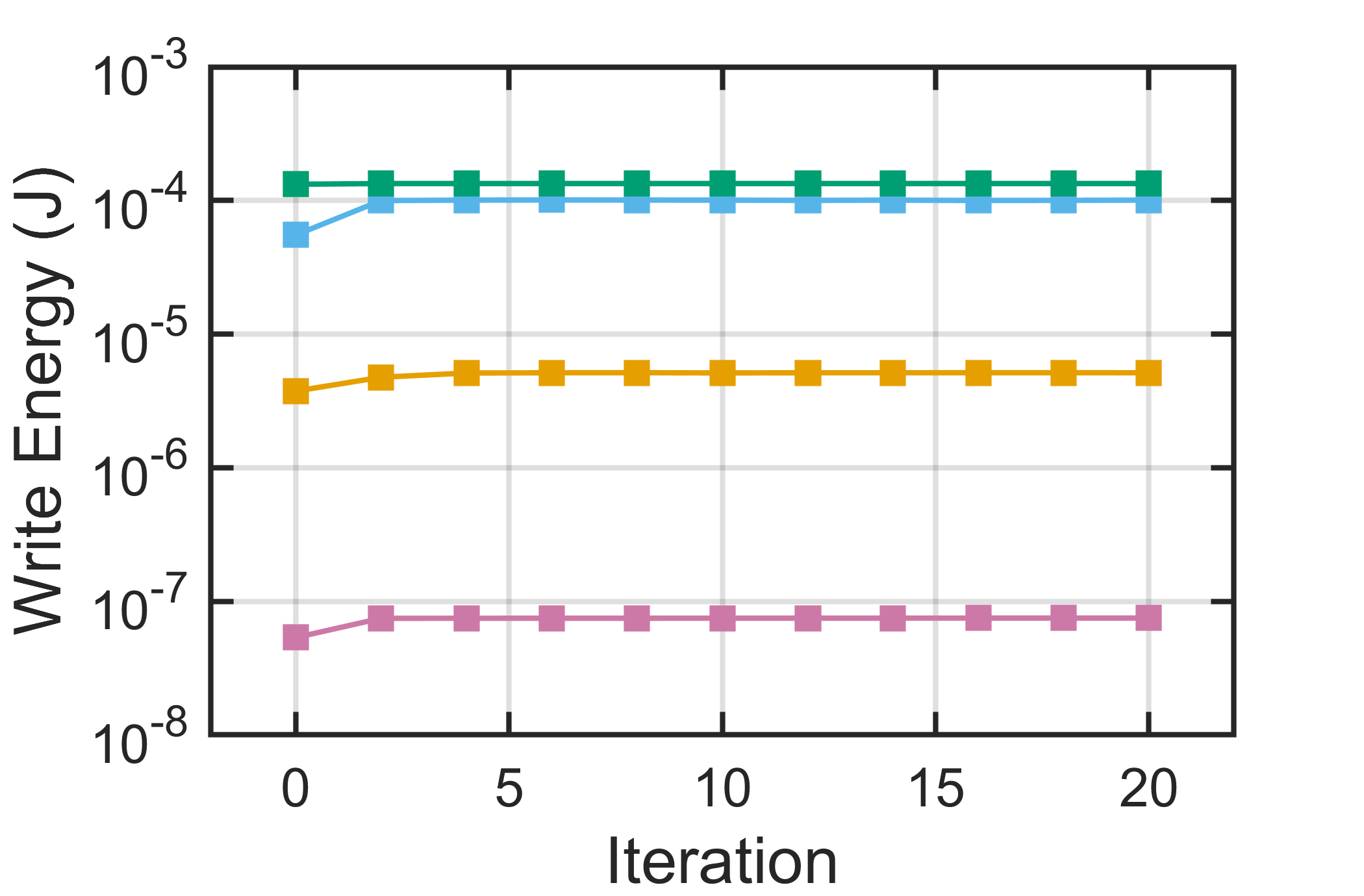}%
        }
    \end{minipage}
    \hfill
    \begin{minipage}[t]{0.49\textwidth}
        \centering
        \subfigure[]{%
            \includegraphics[width=\textwidth]{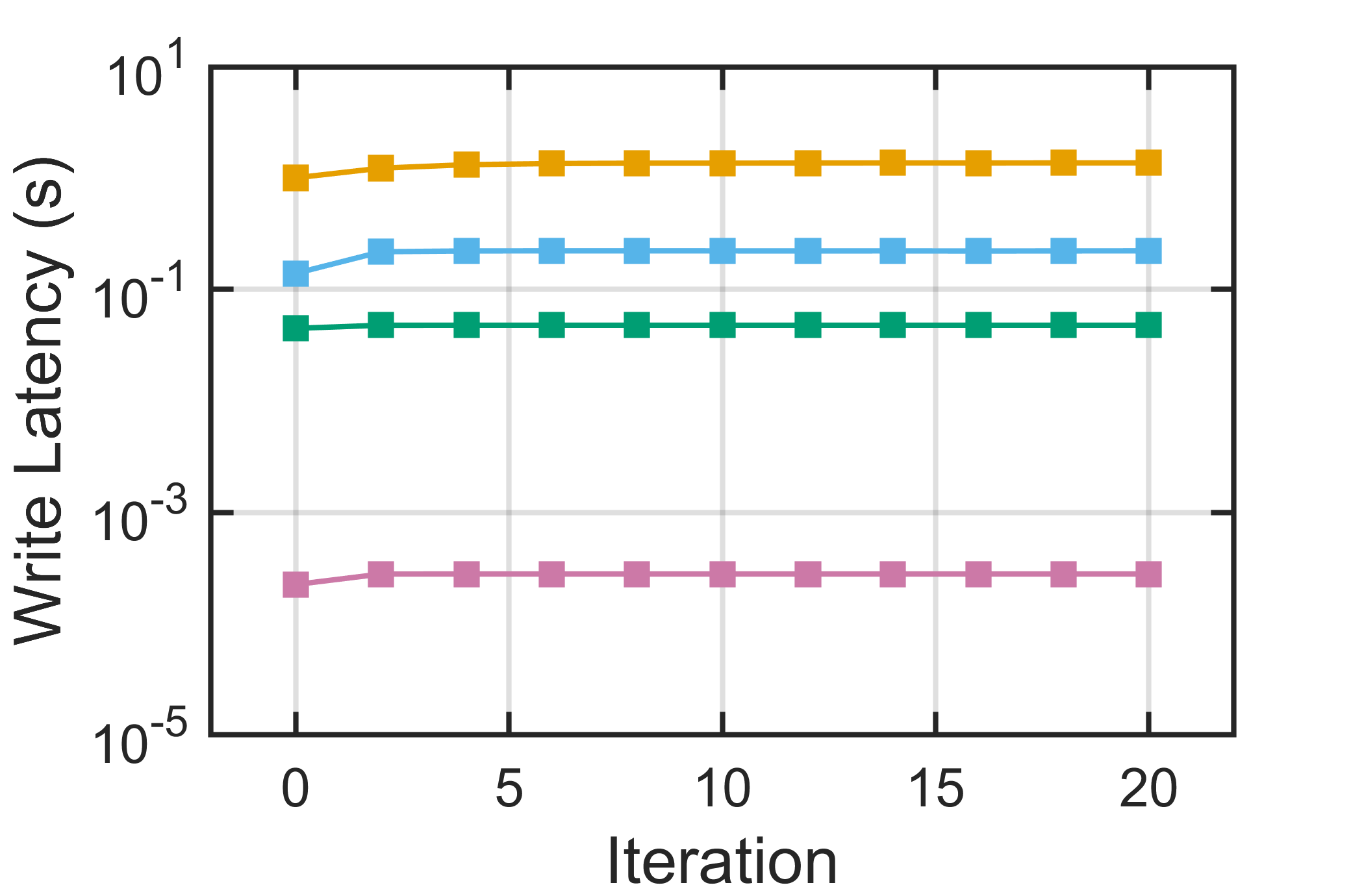}%
        }
    \end{minipage}
    
    \caption{Effects of the~\texttt{adjustableWriteandVerify} function (with no error correction algorithms applied) on the \textbf{bcsstk02} matrix with different fixed numbers of iteration counts: (a) Relative $\ell_2$-norm Error, (b) Relative $\ell_\infty$-norm Error, (c) Write Energy, (d) Write Latency.}
    \label{fig:bcsstk02_EC_0}
\end{figure*}

\begin{figure*}
    \makeatletter
    \renewcommand{\thefigure}{S\@arabic\c@figure}
    \makeatother
    
    \centering
    \begin{minipage}[t]{0.49\textwidth}
        \centering
        \subfigure[]{%
            \includegraphics[width=\textwidth]{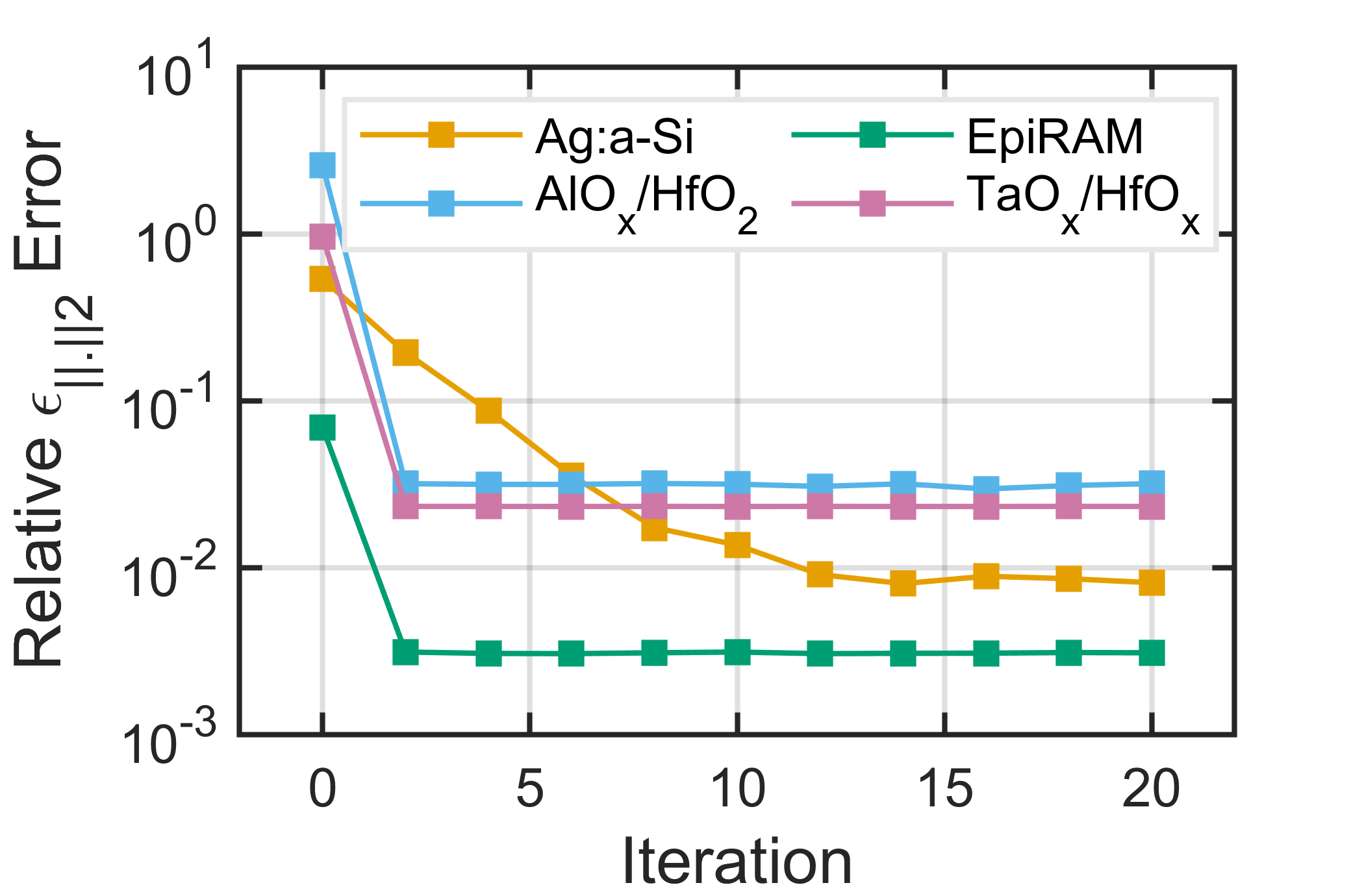}%
        }
    \end{minipage}
    \hfill
    \begin{minipage}[t]{0.49\textwidth}
        \centering
        \subfigure[]{%
            \includegraphics[width=\textwidth]{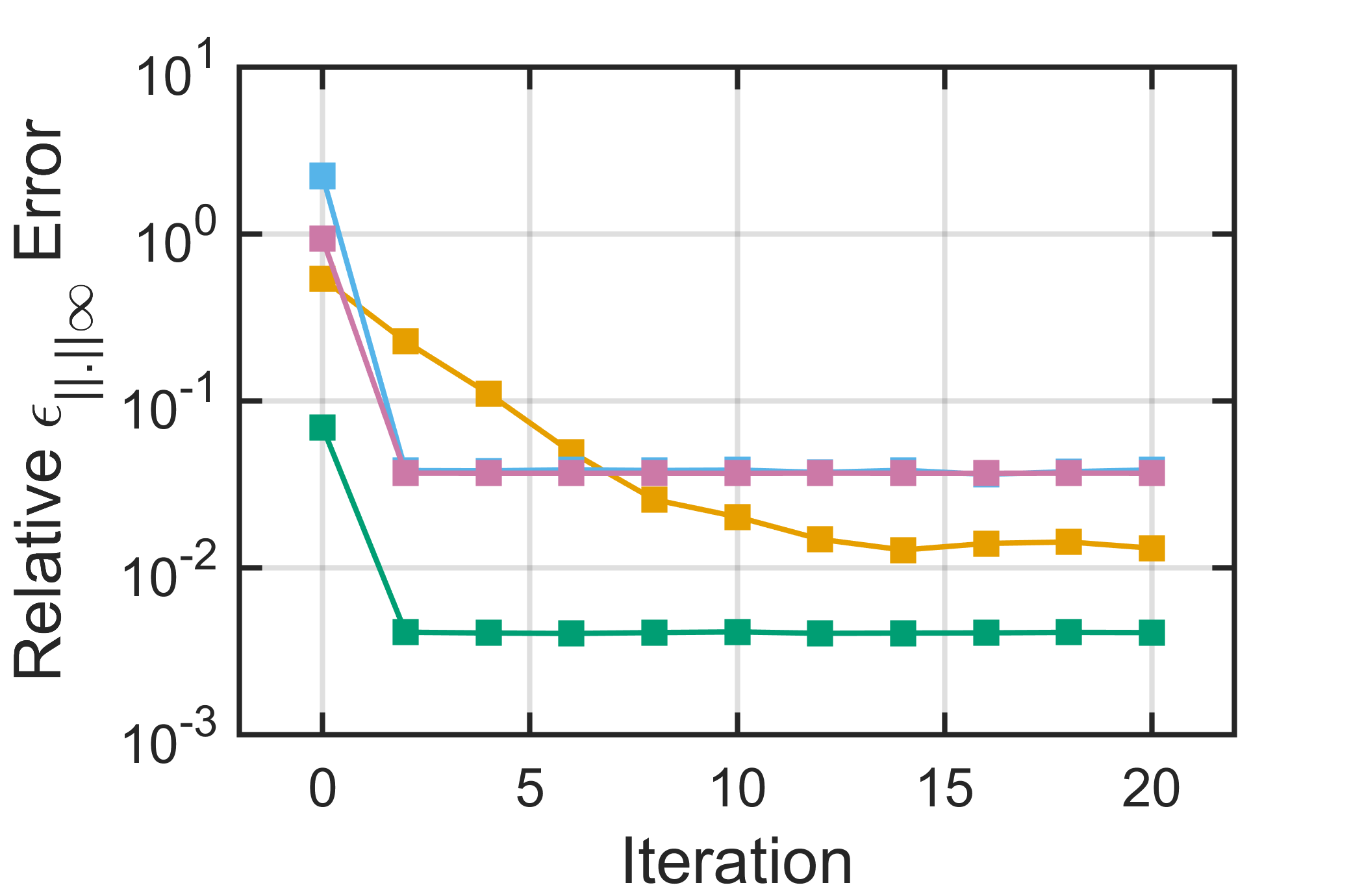}%
        }
    \end{minipage}
    
    \begin{minipage}[t]{0.49\textwidth}
        \centering
        \subfigure[]{%
            \includegraphics[width=\textwidth]{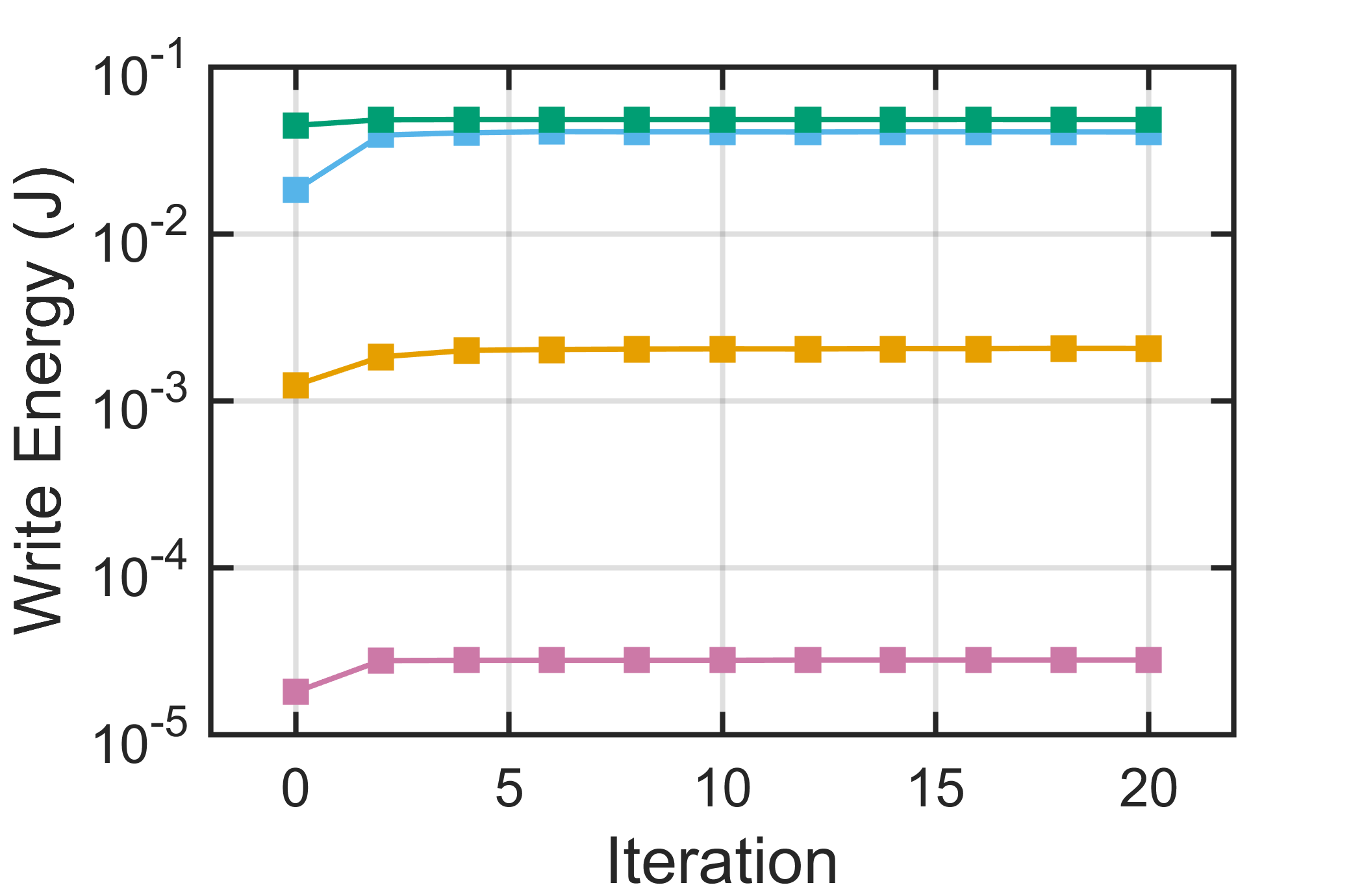}%
        }
    \end{minipage}
    \hfill
    \begin{minipage}[t]{0.49\textwidth}
        \centering
        \subfigure[]{%
            \includegraphics[width=\textwidth]{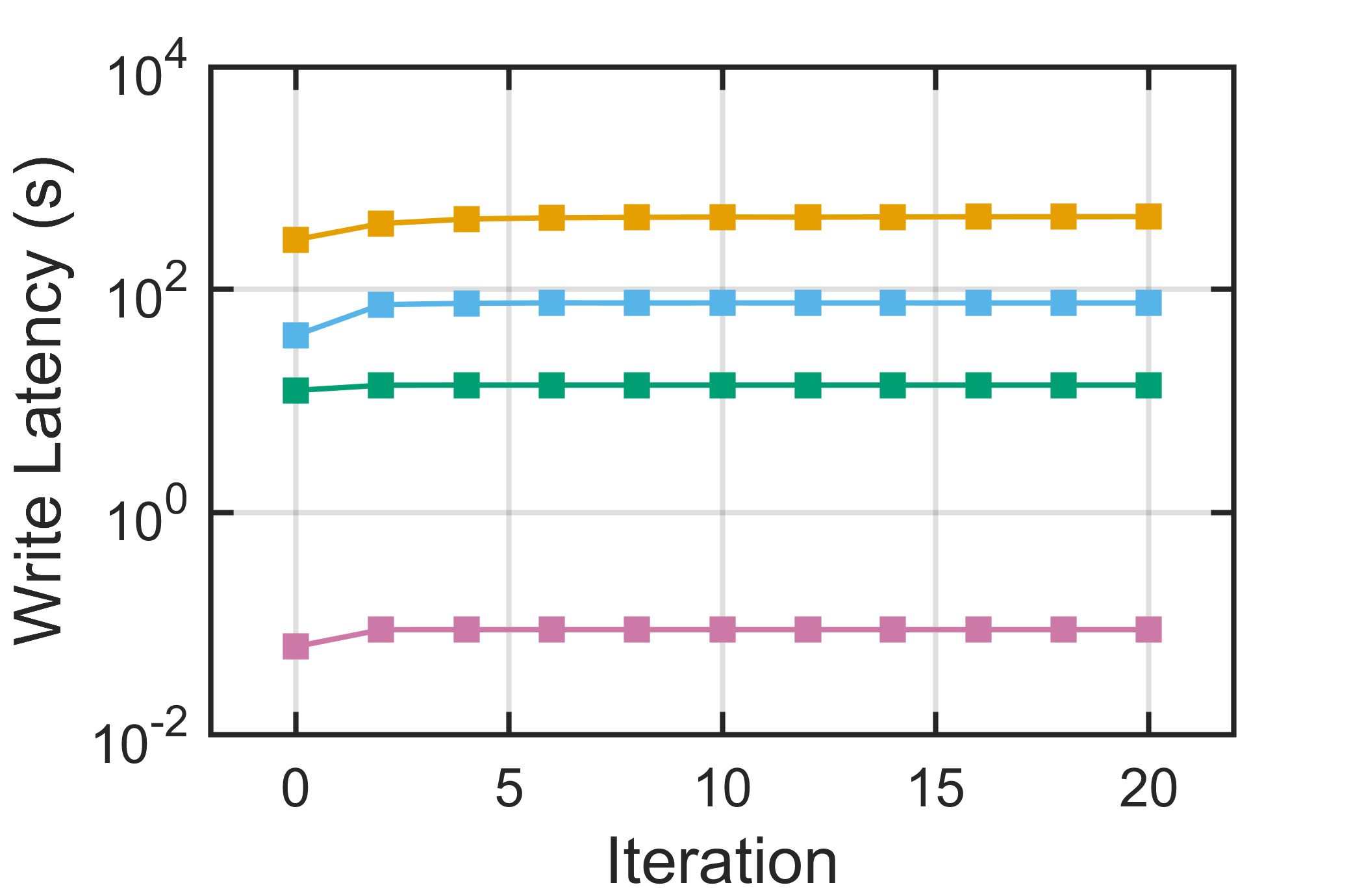}%
        }
    \end{minipage}
    
    \caption{Effects of the~\texttt{adjustableWriteandVerify} function (with error correction algorithms applied) on the \textbf{bcsstk02} matrix with different fixed numbers of iteration counts: (a) Relative $\ell_2$-norm Error, (b) Relative $\ell_\infty$-norm Error, (c) Write Energy, (d) Write Latency.}
    \label{fig:bcsstk02_EC_1}
\end{figure*}

\section{Elucidating the Algorithms}\label{secB1}

\subsection{Second-order error correction}\label{secB1.1}

Recall that after applying the first-order error correction procedure in the \nameref{sec:Methods} section, we obtain the following vector:

\begin{equation*}
    \begin{aligned}
    \mathbf{v} &= \mathbf{\tilde{A}}\mathbf{x} + \mathbf{A}\mathbf{\tilde{x}} - \mathbf{\tilde{A}}\mathbf{\tilde{x}} \\
    &= \mathbf{A}\mathbf{x}(1 - \pmb{\epsilon}_{\mathbf{A}}\pmb{\epsilon}_{\mathbf{x}})
    \end{aligned}
\end{equation*}

This combination effectively cancels the first-order error terms, leaving only second-order errors—that is, the product $\pmb{\epsilon}_{\mathbf{A}} \pmb{\epsilon}_{\mathbf{x}}$. To minimize the impact of second-order errors, we apply a regularized least-squares denoising method. The denoising method is formulated as the following regularized least-squares problem:

\begin{equation*}
    \min_{\mathbf{p}} \quad \|\mathbf{b} - \mathbf{p}\|_2^2 + \lambda \|\mathbf{L} \mathbf{\mathbf{p}}\|_2^2
\end{equation*}

where $\mathbf{L} \in \mathbb{R}^{n\times n}$ is the first-order differential matrix, defined with 1 on the diagonal, -1 on the superdiagonal, and 0 elsewhere, and $\lambda \in (0,\infty)$ is a regularization parameter. The close-form solution of this problem is given as

\begin{equation*}
    \mathbf{y} (\lambda) = (\mathbf{I}_n + \lambda \mathbf{L}^\top \mathbf{L})^{-1} \mathbf{p}
\end{equation*}

where $\mathbf{I}_n$ is an identity matrix of size $n\times n$. This step preserves the structure of the true solution while attenuating noise—that is, second-order errors. The~\texttt{denoiseLeastSquare} function implements our proposed method by constructing $\mathbf{L}$, computing the inverse $(\mathbf{I}_n + \lambda \mathbf{L}^\top \mathbf{L})^{-1}$, and applying it to $\mathbf{p}$.

\begin{algorithm}
\caption{denoiseLeastSquare}\label{alg:denoiseLeastSquare}
\begin{algorithmic}[1]
\Require $\mathbf{p}\in\mathbb{R}^{n\times1}$; $\lambda\in(0,\infty)$; $h\in\mathbb{Z}^{-}$.
\Ensure $\mathbf{b}\in\mathbb{R}^{n\times1}$, $\mathbf{L}\in\mathbb{Z}^{n\times n}$.
\State $n \leftarrow \text{shape}(\mathbf{p})$;
\State$\mathbf{I} \leftarrow \text{eye}(n)$;
\State$\mathbf{L} \leftarrow 
    \begin{cases}
     l_{ij}=1; & \text{if } i - j = 0\\
     l_{ij}=h; & \text{if } j - i = 1 \\
     l_{ij}=0; & \text{otherwise}
\end{cases}$
\State $\mathbf{b} \leftarrow (\mathbf{I} + \lambda\mathbf{L}^\top\mathbf{L})^{-1}\mathbf{p}$;
\end{algorithmic}
\end{algorithm}

\subsection{Matrix-vector multiplication with error corrections}\label{secB1.2}

To ensure accurate encoding onto RRAM, we employ the proposed iterative write-and-verify processes to mitigate device-to-device and cycle-to-cycle variability. These processes are implemented in the~\texttt{adjustableMatWriteandVerify} and~\texttt{adjustableVecWriteandVerify} functions, as discussed in the \nameref{sec:Methods} section. These processes are essential for the matrix-vector multiplication (MVM) operation, implemented by the following~\texttt{correctedMatVecMul} function. Overall, in this function:

\begin{itemize}
    \item We apply the initial encoding step for both input matrix and vector via~\texttt{adjustableMatWriteandVerify} and~\texttt{adjustableVecWriteandVerify} to reduce initial errors.
    \item We apply both first- and second-order error correction methods to obtain a corrected product.
\end{itemize}

\begin{algorithm}
\caption{correctedMatVecMul}\label{alg:correctedMatVecMul}
\begin{algorithmic}[1]
\Require $\mathbf{A}\in\mathbb{R}^{m\times n}$; $\mathbf{x}\in\mathbb{R}^{n}$; $\lambda\in(0,1)$; $h\in\mathbb{Z}^{-1}$; $\pmb{\epsilon}\in\mathbb{R}$; $N\in\mathbb{Z}^+$; $p\in\{2,\infty\}$
\Ensure $\mathbf{b}\in\mathbb{R}^{n}$
\State $\mathbf{\tilde{A}}$ $\leftarrow$~\texttt{adjustableMatWriteandVerify}($\mathbf{A}, \pmb{\epsilon}, N, p$);
\State $\mathbf{\tilde{x}}$ $\leftarrow$~\texttt{adjustableVecWriteandVerify}($\mathbf{x}, \pmb{\epsilon}, N, p$);
\State $\mathbf{\tilde{v}} \leftarrow \mathbf{\tilde{A}}\mathbf{\tilde{x}}$;
\State $\mathbf{\tilde{y}} \leftarrow \mathbf{\tilde{A}}\mathbf{x}$;
\State $\mathbf{\tilde{u}}$ $\leftarrow$ $\mathbf{A}\mathbf{\tilde{x}}$;
\State $\mathbf{p} \leftarrow \mathbf{\tilde{y}} - \mathbf{\tilde{v}} + \mathbf{\tilde{u}}$;
\State $\mathbf{b} \rightarrow \text{denoiseLeastSquare}(\mathbf{p}, \lambda, h)$;
\end{algorithmic}
\end{algorithm}

\subsection{Auxiliary functions for distributed computing}\label{secB1.3}

For large-scale matrices exceeding the physical limits of an MCA system, a distributed paradigm is implemented, enabling scalability through virtualization and parallel processing. To support the distributed paradigm, implemented by the~\texttt{distributedMatVecMul} function in the \nameref{sec:Methods} section, we introduce several auxiliary functions to help facilitate the process. Specifically,

\begin{itemize}
    \item The~\texttt{zeroPadding} function ensures dimension matching between the problem dimension and the system's physical dimension, if necessary, by adding zero rows and columns.
    
        \begin{algorithm}[h]
        \caption{zeroPadding}\label{alg:zeroPadding}
            \begin{algorithmic}[1]
                \Require $m,n,R,C,r,c\in\mathbb{Z}^+$
                \If{$m < R\times r$}: \\
                    \quad $\texttt{setZeroWeightsRows}(R\times r - m)$;
                \EndIf
                \If{$n < C\times c$} \\
                    \quad $\texttt{setZeroWeightsColumns}(C\times c - n)$;
                \EndIf
            \end{algorithmic}
    \end{algorithm}

    \item The~\texttt{generateMatChunksSet} partitions the input matrix $\mathbf{A}$ or a block obtained from~\texttt{blockPartition} (described in the \nameref{sec:Methods} section), into $R \times C$ chunks for individual MCA devices. For large matrices, it first applies~\texttt{blockPartition}, then indexes each block into chunks. For smaller matrices (ideal or non-ideal cases, where dimensions are less than system capacity), it directly indexes $\mathbf{A}$ into chunks. As a result, we ensure compatibility with the multi-MCA configuration, where each MCA processes a chunk in parallel.
    
    \begin{algorithm}[h]
        \caption{generateMatChunksSet}\label{alg:generateMatChunksSet}
        \begin{algorithmic}[1]
        \Require $\mathbf{A}\in\mathbb{R}^{m\times n}; R, r,C,c \in\mathbb{Z}^{+}$. 
        \Ensure A collection of matrix chunks: $\mathcal{S}_{\mathbf{A}}$.
        \State $\mathcal{S}_{\mathbf{A}}\leftarrow \{\emptyset\}$;
    
        \If{$(m > R\times r)$ and $(n > C\times c)$}: 
            \State $\mathcal{B}_{\mathbf{A}} \leftarrow~\texttt{blockPartition}(\mathbf{A}, R,r,C,c)$;
            \For{each block $\mathbf{\hat{A}}_k \in \mathcal{B}_{\mathbf{A}}$}
                    \For{$1\leq p \leq R$}
                        \For{$1\leq q \leq C$}
                            \State $\mathbf{A}_{[p,q]} \leftarrow~\texttt{indexing}(\mathbf{\hat{A}}_k)$;
                            \State $\mathcal{S}_{\mathbf{A}} \leftarrow \mathcal{S}_{\mathbf{A}} \cup \{\mathbf{A}_{[p,q]}\}$
                        \EndFor
                    \EndFor
            \EndFor
    
        \Else
            \For{$1\leq p \leq R$}
                \For{$1\leq q \leq C$}
                    \State $\mathbf{A}_{[p,q]} \leftarrow~\texttt{indexing}(\mathbf{A})$;
                    \State $\mathcal{S}_{\mathbf{A}} \leftarrow \mathcal{S}_{\mathbf{A}} \cup \{\mathbf{A}_{[p,q]}\}$
                \EndFor
            \EndFor
        \EndIf
        \end{algorithmic}
    \end{algorithm}

    \item In a similar manner, the~\texttt{generateVecChunksSet} function partitions the input vector $\mathbf{x}$ into chunks corresponding to matrix chunks, using the ~\texttt{vecBlockPartitioning} protocol for alignment with distributed computation. It ensures that the vector is appropriately divided to match the partitioned matrix chunks, facilitating parallel processing.

    \begin{algorithm}[h]
        \caption{generateVecChunksSet}\label{alg:generateVecChunksSet}
        \begin{algorithmic}[1]
        \Require $\mathbf{x}\in\mathbb{R}^{n}$; $R\in\mathbb{Z}^+$.
        \Ensure A collection of vector chunks: $\mathcal{S}_{\mathbf{x}}$
        \State $\mathbf{x}_{\text{chunk}} \leftarrow~\texttt{vecBlockPartitioning}(\mathbf{x})$
        \State $\mathcal{S}_{\mathbf{x}} \leftarrow \mathcal{S}_{\mathbf{x}} \cup \{\mathbf{x}_{\text{chunk}}\}$
        \end{algorithmic}
    \end{algorithm}
    
\end{itemize}





\newpage

\end{document}